\documentclass[11pt]{JHEP3}
\usepackage{epsfig}
\usepackage{amsfonts}
\usepackage{latexsym}
\usepackage{amsmath}
\usepackage{datetime}
\usepackage{mathrsfs}
\usepackage[square, comma, sort&compress,numbers]{natbib}
\def\half{{1\over 2}}
\numberwithin{equation}{section}

 \def\r{\rightarrow}

\newcommand{\bi}{\begin{itemize}}
\newcommand{\ei}{\end{itemize}}
\newcommand{\non}{\nonumber}
\def\p{\partial}
\def\a{\alpha}
\def\b{\beta}
\def\d{\delta}
\def\g{\gamma}
\def\l{\lambda}
\def\L{\Lambda}
\def\S{\Sigma}
\def\e{\epsilon}

\def\M{\mathcal{M}}
\def\V{\mathcal{V}}
\def\N{\mathcal{N}}
\def\A{\mathcal{A}}
\def\B{\mathcal{B}}
\def\C{\mathcal{C}}
\def\th{\theta}

\def\om{\omega}
\def\Om{\Omega}
\def\s{\sigma}
\def\r{\rightarrow}
\def\half{{\frac12}}
\def\R{{\Bbb  R}}
\newcommand{\bea}{\begin{eqnarray}}
\newcommand{\eea}{\end{eqnarray}}
\newcommand{\be}{\begin{equation}}
\newcommand{\ee}{\end{equation}}

\title{General black holes, untwisted}

\maketitle
\preprint{UPR-1248-T }
\author{\large Mirjam Cveti\v c$^{\dag,\ddag}$, Monica  Guica$^\dag$ and Zain H. Saleem$^{\flat,\dag}$  

\vspace{0.5cm}

{\it $^\dag$ \normalsize{Department of Physics and Astronomy, University of Pennsylvania,}   \vspace{-2 mm} \\

\hspace{-0.3 cm} \normalsize{Philadelphia, PA 19104-6396, USA}
}

 \vspace{ 3 mm}

{\it $^\ddag$ \normalsize{Center for Applied Mathematics and Theoretical Physics},  \vspace{-2 mm}\\
 \hspace{-0.3 cm} \normalsize{University of Maribor, Maribor, Slovenia}
 }

 \vspace{ 3 mm}

{\it $^\flat$ \hspace{-3 mm} \normalsize{ National Center for
Physics , Quaid-e-Azam University Campus, } \vspace{-2 mm} \\
 \hspace{-0.3 cm} \normalsize{Shahdara Valley road, Islamabad, 
Pakistan}
} }

\abstract{

\vskip 5 mm

We use solution-generating techniques to construct interpolating geometries between general asymptotically flat, charged, rotating, non-extremal black holes in four and five dimensions and their subtracted geometries. In the four-dimensional case, this is achieved by the use of Harrison transformations, whereas in the five-dimensional case we use STU transformations. We also give the interpretation of these solution-generating transformations in terms of string (pseudo)-dualities, showing that they correspond to combinations of T-dualities and Melvin twists. Upon uplift to one dimension higher, these dualities  allow us to ``untwist'' general black holes to $AdS_3$ times a sphere. 

}

\begin{document}

\newpage
\section{Introduction}

A baffling property of  general multi-charged rotating  black holes  
in four \cite{CYI} and five  \cite{CYII} dimensions  is that their thermodynamic features, such as the  entropy formula \cite{CYI} and the first law of thermodynamics \cite{Larsen, CL97I, CL97II}, are  strongly suggestive  of a possible microscopic interpretation in terms of a two-dimensional conformal field theory (CFT).  Furthermore, the wave equation for massless scalars in  non-extremal black hole backgrounds  exhibits an approximate $SL(2,\mathbb{R}) \times SL(2,\mathbb{R})$ conformal symmetry at low energies, which is spontaneously broken by the temperatures \cite{CL97I, CL97II,Castro:2010fd}. Thus, one may expect that at least the low-energy dynamics of general black holes are described by a CFT. 

Recently, \cite{CL11I,CL11II} advanced a concrete proposal -  deemed  ``subtraction'' - for how  to  relate general black holes to CFTs. The subtraction procedure  consists of  removing certain terms in the warp factor of the black hole geometry,
in such a way that the scalar wave equation acquires a manifest local $SL(2,\R)\times SL(2,\R)$ conformal symmetry.
The horizon area and the periodicities of the angular and time coordinates remain fixed. For this reason, the subtraction process is expected to preserve the internal structure of the black hole. Given that the  geometry becomes  asymptotically conical \cite{conic}, rather than asymptotically flat,
 the physical interpretation of  subtraction is 
the removal of the  ambient asymptotically Minkowski space-time in a 
way that extracts the ``intrinsic''  $SL(2,\R)\times SL(2,\R)$ 
symmetry of the black hole.

 The subtraction procedure has been explicitly implemented both for   five-dimensional three-charge rotating black holes \cite{CL11I} and four-dimensional four-charge ones \cite{CL11II}.   It works particularly well in the context of the four-dimensional STU model \cite{Duff:1995sm} and its five-dimensional uplift, since in these cases the non-trivial matter field configurations that support the subtracted geometries are still solutions of the same Lagrangian as the original black holes. Moreover, one can use the extra dimensions available in the string-theory embedding of these models to show that the four-dimensional subtracted geometries uplift to  $AdS_3\times S^2$ \cite{CL11II} and the five-dimensional ones, to  $AdS_3\times S^3$ \cite{CL11I}, thus making the connection  with two-dimensional CFTs entirely explicit. 

Nevertheless, the precise relationship between the CFT dual to the (uplifted) subtracted geometry and the microscopic dual of the original, asymptotically flat black hole remains unclear. Building upon the observation of \cite{conic} that the subtracted geometries can be obtained as a particular scaling, low energy limit of the original black holes, \cite{deb} proposed that the latter can be understood by turning on certain irrelevant deformations in the dual CFT. Concretely, for the specific case of static charged black holes, they constructed a set of interpolating solutions between the original  and  subtracted geometries\footnote{For works that studied the interpolating flows for extremal black holes and their relationship to the attractor mechanism  please see \cite{Chakraborty:2012nu,Chakraborty:2012fx}.}. Upon dimensional uplift, these can be interpreted as flows initiated by a set of irrelevant $(2,2)$ operators from $AdS_3 \times S^2$ to the uplift of the original black hole. 

The authors of \cite{deb} argued that even though the deformations are irrelevant, they are nevertheless predictive in the regime of large charges. However, this perspective does not explain
 - especially when the charges are small and the deformations large - 
why the  Cardy-like formula for the black hole entropy holds . More structure is likely present.

An example that may be related is that of dipole theories \cite{berggan,dasgupta,Bergman:2001rw,stromdipwei}, which have been argued in \cite{kerrdip} to be relevant for understanding the microscopic origin of the entropy of extremal black holes \cite{kerrcft}. These theories \emph{can} be understood as irrelevant deformations of a CFT$_2$ by a $(1,2)$ operator; this explains for example the momentum dependence of the conformal dimensions of primary operators \cite{nr}. Nevertheless, the irrelevant deformation picture \emph{does not} explain why the entropy of dipole black holes is described by a Cardy-like formula. Rather, this property of the entropy follows from the fact that dipole backgrounds are related by string pseudo-dualities to backgrounds described by a CFT. These induce a star-product deformation of the gauge theory dual to the background, which is in turn known to not affect the leading entropy. Thus, both complementary points of view are needed to understand the properties of dipole black holes.

We believe that a similar story should hold in the case of subtracted geometries. In fact, it was shown in \cite{conic}  for the case of a Schwarzschild black hole - and in \cite{vir} for the general four-dimensional case - that the original black hole and its subtracted geometry are connected by  certain  ``infinite boost'' Harrison transformations\footnote{Harrison transformations are particular solution generating techniques in the STU model reduced to three dimensions along time. Restricted types of Harrison transformations can act within  Einstein-Maxwell-(dilaton) gravity only, and have recently been employed to study the physics - and especially the ergoregion -  of black holes immersed in strong magnetic fields \cite{Gibbons:2013yq,Yazadjiev:2013hxa}.}. While Harrison transformations do not quite leave the black hole's entropy invariant (a notable exception is the  Kerr black hole), they change it in a controlled way that preserves its Cardy-like form.

The goal of this paper is to further elucidate the relationship between  general black holes and their subtracted geometries. We use solution generating techniques to produce a set of interpolating backgrounds between the two endpoints of interest. 
 In the four-dimensional case, the interpolating  solutions are obtained via Harrison transformations, whereas in the five-dimensional case we use STU transformations, which are closely related to the generalized spectral flows of \cite{Bena:2008wt}. Using these techniques, we are able to reproduce the static interpolating geometries found in \cite{deb} and  to  generalize them to the rotating case and to five dimensions.

It is natural to ask what is the string-theoretical interpretation of the above transformations. Interestingly, in the \emph{static} case, this question has been answered a long time ago by Sfetsos and Skenderis \cite{skendy}, who - building upon previous work \cite{Hyun:1997jv,Boonstra:1997dy} -  showed that all non-rotating charged non-extremal four and five-dimensional black holes in string theory can be related via pseudo-dualities to  $AdS_3 \times S^2$ and  $AdS_3 \times S^3$, respectively. In the case of five-dimensional black holes, the pseudo-dualities that \cite{skendy} employ agree with the duality interpretation of STU transformations given in \cite{stubena}.

 In this article, we extend the results of \cite{skendy} to the general charged rotating case. By slightly modifying their prescription, we are able to relate the subtraction procedure for both four- and five-dimensional black holes to a relatively simple combination of T-dualities and \emph{timelike} Melvin twists. The Melvin twists do not preserve the coordinate periodicities, and thus the transformations we use are strictly pseudo-dualities.  As in the dipole case mentioned above, each block combination  of T-dualities and Melvin twists corresponds to turning on a particular (set of) irrelevant operator(s) in the dual CFT. Thus, every non-extremal rotating black hole can be ``untwisted'' to $AdS_3$ times a sphere via  the above transformations.


%




This paper is organised as follows. In section \ref{stubhsg}, we review the subtraction procedure for four-dimensional STU black holes, and Harrison  transformations in \ref{secsolgen}. In section \ref{disc5du} we show that by applying only three of the four possible Harrison transformations, we already obtain a geometry that uplifts to $AdS_3 \times S^2$, and that the effect of the fourth Harrison is that of a coordinate transformation in five dimensions. In \ref{dualint}, we discuss the duality interpretation of the remaining Harrison transformations.

In section \ref{5dbhtw}, we repeat the procedure for the case of five-dimensional black holes. Many technical aspects and explicit solutions are relegated to the appendices. In particular, appendix \ref{intkerr} contains the five-dimensional uplift of the interpolating geometries for the case of the Kerr black hole. The interpolating solutions for the case of five-dimensional general black holes can be found in  appendix \ref{5dbhint}.

\section{Un-twisting general $4d$ black holes \label{untw4dbh}}

\subsection{STU black holes and subtracted geometries \label{stubhsg}}

In this section, we will be working in the context of the four-dimensional STU model \cite{Duff:1995sm} - an $\N=2$ supergravity theory coupled to three vector multiplets, characterized by the  prepotential\footnote{Throughout this article, we will be using the conventions and definitions of \cite{vir}.}

\be
\mathcal{F} = - \frac{X^1 X^2 X^3}{X^0} \label{prep}
\ee
As usual, the bosonic content of this theory consists of the metric, four gauge fields $A^\Lambda$, $\Lambda = \{ 0,\ldots ,3\}$ and three complex scalars

\be
z^I = \frac{X^I}{X^0} \;, \;\;\;\;\;\; I = \{1,2,3\}
\ee
All the couplings of the theory, as well as the relationship between the various fields are entirely determined by the above $\N=2$ prepotential. 

We consider non-extremal rotating black hole solutions of this theory that are magnetically charged under three of the field strengths, with charges $p^I$, and electrically charged under the fourth field strength, with charge $q_0$. The metric of these solutions can be parametrized as 

\be
ds^2 = -e^{2U} (dt+ \om_3)^2 + e^{-2U} ds_3^2 \label{4dmet}
\ee
The three-dimensional base metric only depends on the rotation $(a)$ and mass $(m)$ parameters of the solutions, and takes the form

\be
ds_3^2 = \frac{G}{X} dr^2 + G d\th^2 + X \sin^2\th d\phi^2 
\ee 

\be
X = r^2 - 2 m r + a^2 \;, \;\;\;\;\; G = r^2 - 2 m r + a^2 \cos^2 \th
\ee
The dependence on the charges is encoded in the conformal factor $U$ and the angular velocity $\om_3$, as well as in the gauge fields and scalars that support the geometry. Parameterizing the charges as 

\be
q_0 = m \sinh 2 \d_0 \;, \;\;\;\;\; p^I = m \sinh 2 \d_I \label{chdef}
\ee
and introducing the shorthands $c_i = \cosh \d_i$, $s_i = \sinh \d_i$, one finds that

\be
\om_3 = \frac{2 m a \sin^2\th}{G} [(\Pi_c-\Pi_s) r + 2 m \Pi_s] d\phi \label{eqom3}
\ee
where

\be
\Pi_c = c_0 c_1 c_2 c_3 \;, \;\;\;\;\;\; \Pi_s = s_0 s_1 s_2 s_3
\ee
The conformal factor $U$ is traded for a new quantity $\Delta$

\be
\Delta \equiv G^2 e^{-4U} \label{defdelta}
\ee
which has the nice property that it is  polynomial in $r$. For the above, asymptotically flat, solutions 

\bea 
\Delta &=& \left(a^2 \cos^2\th + (r+ 2m s_0^2)(r+ 2m s_1^2)\right)\,  \left(a^2 \cos^2\th + (r+ 2m s_2^2)(r+ 2m s_3^2)\right)- \non  \\ 
&& \hspace{2 cm}- 4 a^2 m^2 (s_0 s_1 c_2 c_3 - c_0 c_1 s_2 s_3)^2 \cos^2 \th \label{delta4ch}
\eea
The black hole solutions are also supported by non-trivial gauge fields and scalars\footnote{The scalar and vector sources are related by a subset of U-duality transformations to the original four-charge solution \cite{CYII,CCLP}. The metric is the same.}, whose explicit form can be inferred from the formulae in appendix \ref{gen4chbh}. 

\bigskip

An interesting property of general black holes is that the wave equation for massless scalar perturbations is separable, and moreover it has a low-energy approximate $SL(2,\mathbb{R}) \times SL(2,\mathbb{R})$ symmetry. In order to render this $SL(2,\mathbb{R}) \times SL(2,\mathbb{R})$ symmetry  exact, \cite{CL11II} have introduced the so-called ``subtracted'' geometries, which differ from the original black hole metrics only by a change in the conformal factor $\Delta$

\be
\Delta \r \Delta_{sub} = (2m)^3 r (\Pi_c^2 - \Pi_s^2) + (2m)^4 \Pi_s^2 - (2m)^2 (\Pi_c-\Pi_s)^2 a^2 \cos^2 \th
\ee
The rotation parameter  $\om_3$ in \eqref{eqom3} is kept fixed. Since the asymptotic behaviour of $\Delta_{sub}$ is linear in $r$ - as opposed to quartic - the new solutions are no longer asymptotically flat. Rather, they are asymptotically  Lifshitz with dynamical exponent $z=2$ and  hyperscaling violating exponent $\theta=-2$ (for definition and applications, see e.g. \cite{Dong:2012se}). The physical picture that lies behind this replacement is that the subtraction procedure corresponds to enclosing the black hole into an ``asymptotically conical box'', which isolates its intrinsic dynamics from that of the ambient spacetime, while preserving its thermodynamic properties. 

In \cite{CL11II} it was shown that in the static case the matter fields $A^\Lambda, z^I$ supporting the subtracted geometry are still solutions of the STU model, albeit with unusual asymptotics. Furthermore,  the explicit sources for  the subtracting  geometry of multi-charged rotating black holes were obtained 
 in \cite{conic} as a  scaling limit of certain STU black holes.   Uplifting the subtracted geometries to five dimensions, one finds $AdS_3 \times S^2$ \cite{CL11II}, which realizes the conformal symmetry of the four-dimensional wave equation in a linear fashion. In the following we will try to better understand the relationship between the original, asymptotically flat black holes, their subtracted geometries, and their  five-dimensional uplift.

\subsection{Solution-generating transformations \label{secsolgen}}

A powerful tool that we will be using extensively  are solution-generating transformations that relate backgrounds of the four-dimensional STU model with a timelike isometry. These solution generating techniques can be used to both generate all the charged black holes of the previous subsection from the non-extremal Kerr solution\footnote{ They were employed in \cite{CYII, CCLP} to generate four charge rotating black holes in four dimension with two magnetic and two electric charges. Here we are interested in the solution with one electric and three magnetic charges. }, and to relate these general asymptotically flat black holes to their subtracted geometries.

The procedure is as follows. The four-dimensional STU Lagrangian itself has an
$O(2,2)\sim  SL(2,\R)\times SL(2,\R)$ T-duality symmetry, which is enlarged at the level of the equations of motion to include a third $SL(2,\R)$ 
electric/magnetic S-duality symmetry.  Upon reduction to 
 three dimensions, it is well known \cite{Sen:1994wr} that the ``na\"{i}ve''  $O(3,3)$ three-dimensional  global symmetry is enhanced to
$O(4, 4)$,
 since in three dimensions all one-form potentials can be dualized to scalars.  
 When reducing along time the scalar Lagrangian becomes  a non-linear sigma model  whose
 target space is an $SO(4,4)/(SO(2,2)\times SO(2,2))$  
coset. 


The four-dimensional origin of the sixteen scalars that parametrize the above coset is:


\bi
\item four scalars, $\zeta^\Lambda$, correspond to the electric potentials associated to the vector fields $A^\Lambda$
\item four scalars, $\tilde \zeta_{\Lambda} $, are Hodge dual to the magnetic potentials associated to $A^\Lambda$ 
\item six scalars, $x^I$ and $y^I$, correspond to the real and respectively imaginary parts of the moduli fields $z^I$
\item the scalar $U$ corresponds to $g_{tt}$ in the dimensional reduction \eqref{4dmet}
\item one scalar, $\s$, is Hodge dual to the Kaluza-Klein one-form $\om_3$ 
\ei
The reduction formulae can be found
in  appendix \ref{usff}. The symmetric coset space can be parametrized by the following coset element \cite{Bossard:2009we}

\be
\mathcal{V} = e^{- U \, H_0} \cdot \left(
\prod_{I=1,2,3}
e^{-\frac{1}{2} (\log y^I) H_I} \cdot e^{ - x^I E_I} \right) \cdot
e^{-\zeta^\Lambda E_{q_\Lambda}-  \tilde \zeta_\Lambda E_{p^\Lambda}}\cdot
e^{-\frac{1}{2}\sigma E_0}
\ee
where the $E_{p^\Lambda}$,  $E_{q_\Lambda}$ etc. are generators of the  $so(4,4)$ Lie algebra.
 An explicit parametrization of these generators is given in \cite{vir}. Thus, to any four-dimensional solution of the STU model one can associate a coset element $\mathcal{V}$ via the above procedure.

The $SO(4,4)$ symmetries act simply on the matrix $\M$, defined as

\be
\M = \V^\sharp \V\;, \;\;\;\;\; \V^\sharp = \eta \V^T \eta
\ee
where $\eta$ is the quadratic form preserved by $SO(2,2)\times SO(2,2)$. Namely, if $g \in SO(4,4)$, then the matrix $\M$ transforms as

\be
\M \r g^\sharp \M g
\ee
We will be interested in several specific types of $SO(4,4)$ transformations.

\bigskip

\noindent \emph{\textbf{Charging transformations}}

\medskip

\noindent To each type of electric or magnetic charge that the four-dimensional back hole can have, there is an associated  $so(4,4)$ Lie algebra element that generates it, while leaving the asymptotics of the  solution flat

\be
q_\L \r E_{q_\L} + F_{q_\L} \;, \;\;\;\;\;\; 
p^\L \r E_{p^\L} + F_{p^\L} 
\ee
The expression for the $so(4,4)$ generators $F_{q_\L}$ and $F_{p^\L}$ is again given in \cite{vir}.  Then, the charged black hole discussed in the previous section can be  generated  from the uncharged Kerr black hole by acting with the following group elements 

\be
g_{ch}(q_0,p^I) =   e^{- \d_0 (E_{q_0} + F_{q_0})+\sum_I \d_I (E_{p^I} + F_{p^I})}
\ee
where the various\footnote{Our notation is as follows. The index $I \in \{1,2,3\}$, the \emph{symplectic} index $\Lambda \in \{0,\ldots,3\}$, while the \emph{non-symplectic} index $A \in \{0,\ldots, 3\}$. } $\d_A$ have been defined in \eqref{chdef}. Thus,

\be
\M_{4-charge} = g_{ch}^\sharp \, \M_{Kerr} \,g_{ch}
\ee
In order to obtain the four-dimensional solution, one naturally has to re-dualize the three-dimensional scalars into vectors using \eqref{3dd} and then uplift.

\bigskip

\noindent \emph{\textbf{Rescalings}}

\medskip

\noindent One can also consider the action of the $so(4,4)$ Cartan generators $H_I, H_0$. Letting

\be
g_S = e^{-c_0 H_0 + \sum_I c_I H_I} \label{gscal}
\ee
one finds that they simply rescale the target space scalars as

\be
U \r U + c_0 \;, \;\;\;\;\; \s \r e^{2c_0} \s \;, \;\;\;\;\; x^I \r e^{-2 c_I} x^I \;, \;\;\;\;\; y^I \r e^{-2 c_I} y^I
\ee
\be
\zeta^0 \r  e^{A} \zeta^0 \;,\;\;\;\;\;\zeta^I \r  e^{A-2 c_I} \zeta^I \;, \;\;\;\;\; \tilde \zeta_0 \r e^B \tilde \zeta_0 \;, \;\;\;\;\; \tilde \zeta_I \r e^{B + 2 c_I} \tilde \zeta_I
\ee
where we have let

\be
A= c_0 + \sum_I c_I \;, \;\;\;\;\; B= c_0 - \sum_I c_I 
\ee

\bigskip

\noindent \emph{\textbf{Harrison transformations}}

\medskip

\noindent Harrison transformations are generated by Lie group elements $e^{\a^\L F_{p^\L}}$ or $e^{\tilde \a_\L  F_{q_\L}}$. In this paper, we will only be interested in the following Harrison transformations\footnote{Note that we dropped the tilde on $\a_0$.
}

\be
h_{0} = e^{-\a_0 F_{q_0}} \;, \;\;\;\;\; h_{I} = e^{\a_I F_{p^I}}
\ee
The $h_I$ transformations, eventually accompanied by certain rescalings, have been shown to relate non-rotating black holes to their subtracted geometries in \cite{conic,vir}. In this paper we would like to study the effect of all four Harrison transformation on a given four-dimensional asymptotically flat black hole, carrying arbitrary charge parameters $\d_0, \d_I$. Letting

\be
g_H (\a_0,\a_I) = e^{-\a_0 F_{q_0} + \sum_I \a_I F_{p^I} }
\ee
we compute

\be
\M_{H}(\a_0,\a_I) = g_H^\sharp \, \M_{4-charge} \, g_H  \label{defMH}
\ee 
The effect of the Harrison transformations on the conformal factor $\Delta$ defined in \eqref{defdelta} is to multiply the powers of $r$ by various combinations of $(1-\a_A^2)$, where $A \in \{0,\ldots, 3\}$

\be
\Delta_H = (1-\a_0^2)(1-\a_1^2)(1-\a_2^2) (1-\a_3^2) \, r^4 + \ldots
\ee
in such a way that the coefficient of the $r^4$ term vanishes when any of the $\a_A$ equals one, the coefficient of $r^3$ is zero when any two of the $\a_A$ equal one, and so on. We give an explicit example of such a $\Delta_H$ in \eqref{expldel}. It is thus clear that by performing at least three Harrison ``infinite boosts'' ($\a =1$), we will obtain the same degree of divergence 
of $\Delta$ with $r$ as the subtracted geometry has. 

\bigskip

\noindent \textbf{The subtracted geometry}

\medskip

\noindent To obtain the subtracted geometry, 
to the $\M_H$ defined in \eqref{defMH}, we need to further apply a scaling transformation of the type \eqref{gscal}. We find that when

\be
\a_I =1 \;, \;\;\;\;\; 
\a_0 = \frac{\Pi_s \cosh \d_0 - \Pi_c \sinh \d_0}{\Pi_c \cosh \d_0 - \Pi_s \sinh \d_0} \label{sola0}
\ee

\be
e^{2c_0} = \frac{e^{\d_1 + \d_2 + \d_3}}{\Pi_c \cosh \d_0 - \Pi_s \sinh \d_0} \;, \;\;\;\;\; e^{2c_I} = \frac{e^{2\d_I}}{2m} \, e^{2 (c_0 -\d_1 - \d_2 - \d_3)} \label{rescc}
\ee
we recover precisely the subtracted geometries of \cite{CL11II} in the general rotating charged case. This result is very similar to that of \cite{vir}, who showed that the subtracted geometry of a general charged rotating black hole can be obtained by applying the $h_I$ Harrison transformations followed by a particular charging transformation and rescalings. We will further comment on the relationship with the result of \cite{vir} at the end of the next subsection.

The set of solutions to the STU model encoded in the matrix $\M_H(\a_0,\a_I)$ represent a four-parameter family of interpolating solutions between the original black hole and its subtracted geometry. In the non-rotating case, these interpolating solutions precisely coincide with those of \cite{deb}, which we review in  appendix \ref{deboer}. We also present the  solution interpolating from the Kerr black hole to its subtracted geometry in appendix \ref{intkerr}.

\subsection{Discussion of the five-dimensional uplift \label{disc5du}}

The microscopic interpretation of the subtracted geometry is clearest in the five-dimensional picture, since its uplift is  $AdS_3 \times S^2$, which is holographically described by a CFT$_2$.  In this subsection we will consider the five-dimensional uplift of a slightly generalized version of the subtracted geometries, namely the Harrison-transformed black holes with $\a_I=1$ and $\a_0$ arbitrary. Interestingly, all these backgrounds uplift to $AdS_3 \times S^2$, irrespective of the value of $\a_0$.

The uplift Ansatz is given by

\be
ds_5^2 = f^2 (dz+A^0)^2 + f^{-1} ds_4^2 \;, \;\;\;\;\; f = (y^1 y^2 y^3)^{\frac{1}{3}}  \label{5dlift}
\ee
where $ds_4^2$ is given in terms of the three-dimensional fields by \eqref{4dmet} and $A^0$ by \eqref{uplifta}.  Plugging in the solution discussed above we obtain

\be
ds^2_5 = ds_3^2 + \ell^2 \left[ d\th^2 + \sin^2\th \left(d\phi - \frac{a e^{-\d_0-\d_1-\d_2-\d_3}}{4 m^2}\, (dt + (\a_0-1) dz)\right)^2 \right] \label{5dsubmet}
\ee
where

\be
\ell= 2 m \, e^{\frac{2}{3}(\d_1 + \d_2 + \d_3)}
\ee
is the radius of the $S^2$. The three-dimensional part of the metric, $ds_3^2$, is $AdS_3$ of radius $2\ell $  in an unusual coordinate system

\bea
ds_3^2 & = & \frac{\ell^2 dr^2}{r^2 - 2 m r +a^2} +  \frac{e^{ - \frac{2}{3} (\d_1+\d_2+\d_3)}}{4m^2} \left[ - ( a^2 e^{-2\d_0} + 2 m r-4m^2 c_0^2 ) \, d \tilde t^2 + \right. \non \\ &&\non \\
& + &  \left.  2(a^2 e^{-2\d_0}  + 4 m^2 c_0 s_0 ) \, d\tilde t dz + (2 m r  -a^2 e^{-2\d_0}   + 4 m^2 s_0^2)\, dz^2\right] \label{uglya0met}
\eea
The entire $\a_0$ dependence is encoded in the new coordinate $\tilde t$

\be
\tilde t = t + \a_0 z
\ee
Thus, the effect of the $h_0$ Harrison transformation, which is non-trivial from a four-dimensional perspective, corresponds to a simple coordinate transformation in five dimensions\footnote{When $\a_0 =1$, the $AdS_3$ factor can be written as a $U(1)$ Hopf fibre over $AdS_2$, where the Hopf fibre coordinate is the fifth dimension $z$. Also, the $z$ component of the Kaluza-Klein gauge field in \eqref{5dsubmet} vanishes.  
Thus, for $\a_0=1$, the four-dimensional geometry itself becomes $AdS_2 \times S^2$. This is in agreement with the well-known result that when all $\a_A$ are equal,  the resulting Harrison transformation acts within   Einstein-Maxwell gravity only, and that 
in the infinite boost  limit it transforms the  Schwarzschild metric  to the Robinson-Bertotti one. This type of transformation was recently employed  in \cite{klemm}.
 }. In  appendix \ref{a0har} we show that the effect of the $\a_0$ Harrison transformation on the five-dimensional uplift of \emph{any} four-dimensional STU geometry with a timelike isometry is that of the  coordinate transformation $t \r t+ \a_0 z$.

The above five-dimensional geometry is supported by magnetic flux through the $S^2$, given by

\be
A^I = e^{2\d_I} \left(\frac{a}{2m} (dt + (\a_0-1) dz) \, e^{-\d_0 - \d_1 -\d_2 -\d_3} - 2 m d\phi \right) \, \cos \th 
\ee
The associated magnetic charges are 

\be
p^I = 2 m e^{2\d_I}
\ee
Note that they are different from the original charges \eqref{chdef}. 
The Brown-Henneaux asymptotic symmetry group analysis \cite{Brown:1986nw} applied to the $AdS_3$ factor \eqref{uglya0met} yields a 
 central charge 

\be
c = \frac{3 (2\ell)}{2 G_3} = \frac{12 \pi \ell^3}{ G_5} 
= \frac{48 m^3}{G_4}\,  e^{2(\d_1+\d_2+\d_3)}
\ee
It is easy to check  that $c= 6\, p^1 p^2 p^3$, as expected. 


\bigskip

Let us now understand the five-dimensional uplift of the subtracted geometry itself. As explained, in order to get precisely the subtracted geometry one needs to perform the additional rescaling transformations $H_0, H_I$, with coefficients given by \eqref{rescc}. From a five-dimensional point of view, these transformations simply multiply the metric by an overall factor 

\be
ds_5^{'2}  = e^{\frac{2}{3} (c_1+c_2+c_3) - 2 c_0} ds_5^2 \label{resmet}
\ee
provided that we replace $t$ and $z$ by the rescaled coordinates 

\be
 t' = e^{2c_0} t \;, \;\;\;\;\; z' = e^{c_0-c_1-c_2-c_3} z \label{rescoo}
\ee
Under the above rescaling, the radius of the $AdS_3$ becomes $\ell_{AdS_3} = 2\sqrt{2 m}$. The associated Brown-Henneaux central charge is then

\be 
c= \frac{6 (2m)^{\frac{3}{2}}}{G_4} \label{cc}
\ee
which only depends on the mass parameter. The action of the rescalings on the magnetic fields is

\be
A^I \r  e^{-c_0 + c_1 + c_2 + c_3 - 2 c_I} A^I \label{resgauge}
\ee
which implies that all magnetic charges are now equal $p^1=p^2=p^3 = \sqrt{2m}$.
One can easily perform a coordinate transformation to put the  metric \eqref{uglya0met} into BTZ form\footnote{The parameters $T_\pm$ are related to the dimensionless left/right moving temperatures in the dual CFT as $T_+ = \pi T_L$, $T_- = \pi T_R$. This redefinition slightly changes the form of  Cardy's entropy formula \eqref{cardyf}.}

\be
\frac{ds^2}{\ell^2} =  T_-^2 du^2 + T_+^2 dv^2 + 2 \rho \, du dv + \frac{d\rho^2}{4(\rho^2-T_+^2 T_-^2)} \label{btzform}
\ee
where we have defined

\be 
u = \frac{\sqrt{m^2-a^2}}{8 m^2 T_-} (-t' + (1+\a_0) z')\, e^{-\sum_I \d_I +\d_0}\;, \;\;\;\;\; v= \frac{1}{8 m T_+} (t' + (1-\a_0) z')\, e^{-\sum_I \d_I-\d_0}
\ee

\be
r = m + \frac{\sqrt{m^2-a^2}}{T_+ T_-} \, \rho
\ee
Requiring that $u,v$ be identified mod $2\pi$ as $z \r z+ 2 \pi$ and plugging in the values \eqref{sola0}, \eqref{rescc} for $\a_0, c_A$ fixes the temperatures to

\be
T_+ = \frac{(\Pi_c-\Pi_s) \sqrt{m}}{2 \sqrt{2}} \;, \;\;\;\;\; T_- = \frac{(\Pi_c+\Pi_s) \sqrt{m^2-a^2}}{2 \sqrt{2 m} }
\ee
It is then trivial to check that  the Cardy formula in the dual CFT 

\be
S_{Cardy} = \frac{\pi}{3} \, c \,(T_+ +T_-) \label{cardyf}
\ee
with $c$ given by \eqref{cc}, reproduces the Bekenstein-Hawking entropy of the general rotating black hole

\be
S_{BH} = \frac{2\pi m}{G_4} [(\Pi_c -\Pi_s) (m+ \sqrt{m^2-a^2}) + 2 m \Pi_s ]
\ee
The central charge \eqref{cc} does not agree with the Kerr/CFT central charge $c= 12 J$ in the extremal limit. This could be explained by the fact that we are using different ``frames'' for computing the entropy. Nevertheless, we can bring the central charge to any desired value while keeping the entropy invariant by performing any rescaling transformation with $c_0 =0$. Under it, the central charge transforms as

\be
c \r c \, e^{c_1+c_2 +c_3}
\ee  
while  the temperatures transform in the opposite way, thus leaving \eqref{cardyf} unchanged. We further discuss these rescalings in the next subsection.

Finally, let us comment on the relationship with \cite{vir}. In that paper, the author applies the three maximal $h_I$ Harrison transformations (followed by certain rescalings) to a black hole with arbitrary magnetic charges $\d_I$, but with electric charge given by  $\tilde \d_0$, where

\be
\sinh \tilde \d_0 = \frac{\Pi_s}{\sqrt{\Pi_c^2-\Pi_s^2}}
\ee
rather than $\d_0$. Also, he does not use the $h_0$ Harrison transformation at all to reach the subtracted geometry.

Of course, one can reinterpret this procedure as starting with a general black hole with charge parameters $\d_0,\d_I$, to which one applies the $h_I$ Harrison transformations with $\a_I=1$, and then performs a charging transformation with parameter $\tilde \d_0 - \d_0$, followed by certain rescalings. It is not hard to check that the $q_0$ charging transformation simply corresponds to a boost in five dimensions. Thus, Virmani's procedure to obtain the subtracted geometry and ours simply differ by a five-dimensional coordinate transformation and some rescalings. Note that in both cases, the parameters of the transformations only depend on $\a_0$, $\Pi_s$ and $ \Pi_c$.

\subsection{Duality interpretation \label{dualint}}

The uplift of the subtracted geometry is $AdS_3 \times S^2$, supported by magnetic fluxes. This is the near-horizon geometry of three intersecting M5-branes in M-theory \cite{Papadopoulos:1996uq,Tseytlin:1996bh}, each of which  wraps a different four-cycle on a six-torus $T^6$

\medskip

\begin{center}
\begin{tabular}{c|ccccccc}
 & $w^1$ & $w^2$ & $w^3$ & $w^4$ & $w^5$ & $w^6$ & $ z$ \\ \hline
M5 &  & & -& -& - & - & - \\
M5 &- &- & & & - & -& - \\
M5 & -& - & - & -& & & - \\
$p$ & &&&&&& - 
\end{tabular}
\end{center}

\noindent Here $z$ denotes the M-theory direction. The number of branes of each type is given by the flux of the corresponding gauge field through the $S^2$. Before the rescalings, we have

\be
p^1= 2 m e^{2\d_1}  \;,\;\;\;\;\; p^2= 2 m e^{2\d_2} \;,\;\;\;\;\; p^3 = 2 m e^{2\d_3} 
\ee
whereas after the scaling transformations we have $p^1=p^2=p^3=  \sqrt{2m}$. The dual CFT (known as the MSW CFT), whose central charge $c=6 p^1 p^2 p^3$ has been microscopically derived in \cite{msw}, describes the low-energy excitations of the M5-brane worldvolume theory.

The above $AdS_3 \times S^2$ geometry has been obtained by applying to the five-dimensional uplift of a general non-extremal rotating four-dimensional black hole a set of $h_I$ Harrison transformations with $\a_I =1$, followed by an $\a_0$ Harrison with parameter \eqref{sola0} and a rescaling. We will analyze the string/M-theory duality interpretation of each of these transformations,  in reverse order.

\subsubsection*{The rescalings}
 
The action of the rescalings \eqref{gscal} on the five-dimensional geometry is given by \eqref{resmet}, \eqref{rescoo} and \eqref{resgauge}. Since they change the radius of $AdS_3$, the M5 magnetic fluxes and the periodicity of the M-theory circle parametrized by $z$, these transformations do not act within the same theory. Rather, they take us from a given MSW CFT to another, of different central charge and defined on a circle of a different radius.

These transformations also do not generally leave the entropy invariant. On the $AdS_3$ length and temperatures they act as

\be
\ell \r e^{\frac{1}{3}(c_1+c_2+c_3) -c_0}\, \ell \;,\;\; \;\;\;\;\; T_\pm \r e^{c_0 - (c_1+c_2+c_3)} \, T_\pm
\ee
Since the central charge $c \propto \ell^3$, they leave invariant Cardy's formula \eqref{cardyf} only if $c_0=0$. It is interesting to note that the only case in which the rescalings are not needed in order to match the entropy is that of the neutral Kerr black hole, which is also the one of most phenomenological interest.

\subsubsection*{The $\a_0$ transformation}

In appendix \ref{a0har}, we show that the $\a_0$ Harrison transformation always corresponds to a coordinate transformation in M-theory, mixing the $AdS_3$ boundary coordinates as 

\be
\left( \begin{array}{c} z \\ t \end{array} \right)\r \left( \begin{array}{cc} 1 &\; 0\; \\ \a_0 & 1 \end{array} \right) \left( \begin{array}{c} z \\ t \end{array}  \right) \label{a0transf}
\ee
Note that the above diffeomorphism is not part of the Brown-Henneaux asymptotic symmetry group, because it mixes the left- and right-moving coordinates $u=z-t$ and $v=z+t$. Thus, this transformation changes the metric on the $AdS_3$ boundary, and therefore it corresponds to turning on a source for the dual stress tensor. For $\a_0$ infinitesimal, we have

\be
S_{CFT} \r S_{CFT} - \a_0 \int dt dz \,T^{zt} 
\ee 
This would be the entire story if the theory was defined on the plane. Nevertheless, in our case the M-theory circle is identified as $z \sim z + 2 \pi$, so the theory is defined on the cylinder. The transformation \eqref{a0transf} \emph{does not}
preserve the cylinder, and thus it is not a symmetry of the theory. In particular, it changes the entropy of the black holes.
 It would be interesting to precisely understand the holographic dual of this coordinate reidentification.

\subsubsection*{The $\a_I$ transformation}

As we have discussed, 
the formula for the conformal factor $\Delta$ is completely symmetric under the interchange of the $\a_A$. In the above, we have shown that the  $h_0$  Harrison transformation corresponds to uplifting to M-theory and performing a specific coordinate transformation. It is then natural to ask whether the remaining $\a_I$ can also be interpreted as coordinate transformations in the appropriate frame. 

That the answer should be yes is rather clear from the work of \cite{skendy}. Those authors showed that a general \emph{static} black hole can be ``untwisted'' to $AdS_3$ by going to the duality frame in which each of its charges becomes momentum and then performing an $SL(2,\mathbb{R})$ transformation in the $(t,z)$ directions. 

The black holes that we are considering carry D0 and D4 charges associated to various four-cycles in the compactification $T^6$. We have already observed that the $h_0$ Harrison transformation corresponds to uplifting to M-theory (which turns the D0 charge into momentum) and then performing 
 the ``shift'' $SL(2,\mathbb{R})$ transformation \eqref{a0transf}. The remaining three Harrison transformations should then be identified with combinations of four T-dualities (which turn a given D4 into D0, and thus M-theory momentum), the shift transformation, reduction  to type IIA, and four T-dualities back. 
In appendix \ref{aitransf} we show that, indeed, these combination of T-dualities and coordinate transformations has the same effect on certain scalars as the corresponding Harrison transformation.



Thus, we have succeeded in  extending the results of \cite{skendy} to general rotating black holes.  While we only considered Harrison transformations represented by matrices of the form \eqref{a0transf}, \cite{skendy} also considered more general $SL(2,\mathbb{R})$ transformations

\be 
\left( \begin{array}{cc} \;a\; &\; b\; \\ c & d \end{array} \right) \;, \;\;\;\;\;\; a d - b c=1
\ee
The entries were further constrained by a condition essentially equivalent to reducing the degree of divergence of $\Delta$. 
It was found that for the specific choice 

\be
 a= \cosh^{-1} \d\;, \;\;\;\; b=0 \;, \;\;\;\;\; c = - e^{-\d}\;, \;\;\;\;\; d= \cosh \d
\ee
the entropy of the black hole is also preserved. As we have already discussed, all transformations with $c \neq 0$ do not preserve the cylinder that the theory is defined on, so they  generically change the entropy, as we saw explicitly in the previous section. It is thus very interesting that - at least in the static case - there exists a choice of $SL(2,\mathbb{R})$ transformations that leave the entropy invariant. It would be instructive to check whether this choice persists in the general rotating case.

\section{Un-twisting $5d$ black holes \label{5dbhtw}}

\subsection{Setup}

Let us now turn to the analysis of five-dimensional black holes. We   consider the non-extremal rotating generalization of the D1-D5-p black hole, first presented in \cite{CYII}.  These black holes are solutions of $\N=2$ $5d$ supergravity coupled to two vector multiplets. The metric can again be written as a timelike fibre over a four-dimensional base space\footnote{In this section we completely reset the notation of the previous one. Thus, the quantities $\Delta, G, \Pi_s, \Pi_c, \ell$ etc. have different interpretation from before. There is no simple relationship between the four-dimensional black holes studied in the previous section and the five-dimensional ones we study now. }

\be
ds_5^2 = - \Delta^{-\frac{2}{3}} \tilde G (dt+\A^t)^2 + \Delta^{\frac{1}{3}} d\hat s_4^2 \label{5ddelta}
\ee
The four-dimensional base space is spanned by the spatial coordinates $\{r,\th, \phi, \psi\}$, and its metric is given by \eqref{dshatsq}. As before, the base metric does not depend on the charges, but only on the mass and rotation parameters\footnote{This suggests that one should be able to generate the general solutions of \cite{CYII}  by only reducing to four dimensions along time, rather than along both time and $\psi$ as it was originally done.}. The remaining quantities are

\be
\Delta = f^3 H_0 H_1 H_5  \;, \;\;\;\;\; \tilde G= f(f-2m)
\ee
where

\be
 H_i = 1+ \frac{2m  \sinh^2\d_i}{f} \;, \;\;\;\;\; f = r^2 + a^2 \cos^2\th + b^2 \sin^2\th \label{deff}
\ee
%
%
As before, the parameters $\d_i$ encode the electric charges of the black hole. As $r \r \infty$, $\Delta \propto r^6$, and the solutions are asymptotically flat. The main observation of \cite{CL11I} was that if one changes the conformal factor $\Delta$ as

\be
\Delta \r \Delta_{sub} = (2m)^2 f (\Pi_c^2-\Pi_s^2) + (2m)^3 \Pi_s^2 \label{deltasub}
\ee
while keeping $\A^t$ and $d\hat s_4^2$ fixed, the wave equation of a massless scalar propagating in the black hole geometry has exact local $SL(2\mathbb{}R) \times SL(2,\mathbb{R})$ symmetry and the black hole thermodynamics is unchanged. In the five-dimensional case, the definition of $\Pi_c$ and $\Pi_s$ has changed to

\be
\Pi_c = c_0 c_1 c_5 \;, \;\;\;\;\; \Pi_s = s_0 s_1 s_5
\ee
Moreover, \cite{CL11I} showed that the five-dimensional subtracted geometry uplifts to $AdS_3 \times S^3$, thus geometrically realizing the hidden conformal symmetry visible in five dimensions. 

In this section we will show that the ``subtraction'' procedure can again be implemented using combinations of string dualities and coordinate transformations. As before, these transformations act naturally in one dimension higher, in this case six dimensions.  Thus, we uplift the metric  to a six-dimensional black string \cite{Cvetic:1998xh} using

\be
ds_6^2 = G_{yy} (dy + A^3)^2 + G_{yy}^{- \frac{1}{3}} ds_5^2 \;, \;\;\;\;\; G_{yy} = \frac{H_0}{\sqrt{H_1 H_5}}
\ee
where the Kaluza-Klein gauge field $A^3$ can be found e.g. in \cite{conic}. This black string is a solution of a very simple six-dimensional theory, namely

\be
S = \int d^6x \sqrt{g} \left( R + (\p\phi)^2 - \frac{1}{12} F_{(3)}^2 \right) \label{6dact}
\ee
which contains a three-form gauge field and a dilaton in addition to the metric. This theory is a consistent truncation of the type IIB supergravity action on $T^4$ with only Ramond-Ramond three-form field. 

Given that the uplifts of  both the original and the subtracted geometry are solutions of the  theory \eqref{6dact} that share the same base metric $d\hat s^2_4$, it is natural that they be related by a symmetry that the six-dimensional action acquires upon reduction to four dimensions along $\{y,t\}$. 
The symmetries of the resulting four-dimensional action are nothing but the STU $SL(2,\mathbb{R})^3$ symmetries.
 The action of STU transformations directly
on the six-dimensional geometry  has been worked out in \cite{stubena}. In the following subsection we will briefly review these transformations and show that they indeed connect the uplifts of the original and subtracted five-dimensional geometries.

\subsection{Subtraction via STU }

STU transformations are the symmetries of the $\mathcal{N}=2$ four-dimensional supergravity theory with prepotential \eqref{prep}. This theory  can be understood  as  the dimensional reduction of the six-dimensional action \eqref{6dact} on a two-torus. From the six-dimensional perspective, the STU transformations relate solutions of \eqref{6dact} which can be written as $T^2$ fibrations over the same four-dimensional base. We parametrize the metric as

\be
ds_6^2 = ds_4^2 + G_{\a\b} (dy^\a + \A^\a) (dy^\b + \A^\b) \;, \;\;\;\;\;\; y^\a = \{y,t\} \label{6dgenmet}
\ee 
The six-dimensional $C^{(2)}$ field can be similarly decomposed as

\be
C^{(2)}_{\a\b} = \zeta \hat \e_{\a\b} \;, \;\;\;\;\; C^{(2)}_{\mu \a} = \mathcal{B}_{\mu \a}- C_{\a\b} \A^\b \non
\ee 

\be
C^{(2)}_{\mu\nu} = \mathcal{C}_{\mu\nu} - \A^\a_{[\mu} \B_{\nu]\a} + \A^\a_\mu  C_{\a\b} \A^\b_\nu \label{decc}
\ee
and there is additionally the dilaton $\phi$. We will be interested in the general rotating black string solution of \cite{Cvetic:1998xh}. We give the expressions for the four-dimensional fields $\A^\a$, $\B_\a$, $G_{\a\b}$, $ ds_4^2$, $\zeta$, $\phi$ that characterize this solution in appendix \ref{genbs}.

Let us now briefly review the interpretation  of the  STU transformations in the type IIB frame, which is discussed at length in \cite{stubena}.  The last one, U, simply corresponds to a coordinate transformation in six dimensions, of the type

\be
\mathcal{U} : \; \left(\begin{array}{c}y   \\ t \end{array} \right) \r  \left(\begin{array}{cc}\; a \; & \; b \; \\ c & d \end{array} \right)\left(\begin{array}{c} y \\ t  \end{array} \right)
\ee
where $ad-bc=1$. 

The T transformation corresponds to a type IIB S-duality, followed by a T-duality along $y$, then by a coordinate transformation as above, a T-duality back on the new $y$ coordinate, and finally an S-duality back. At least when $a=d=1$ and $b=0$, it was shown in \cite{stubena} that it  can  alternatively be interpreted as

\bi
\item a T-duality along $y$
\item  a timelike  Melvin twist with $t \r t + c\, x^{11}$ 
\item  a T-duality back. 
\ei
The S transformation is the same as the T  transformation, both preceded and followed by four T-dualities on the internal $T^4$, whose role is to implement $6d$ electromagnetic duality on the initial and final geometries.

\bigskip

\noindent \emph{\textbf{The T transformation}}

\medskip

\noindent
The first transformation that we will apply to the black string solution \eqref{4dmetstr} - \eqref{gab} is a T-type transformation, given by the $SL(2,\mathbb{R})$ matrix

\be
\mathcal{T} = \left(\begin{array}{cc}\; 1 \; & \; 0 \; \\ \l_1 & 1 \end{array} \right)
\ee
This transformation acts on  \eqref{6dgenmet} as

\be
ds^2_6 \r \sqrt{\S_1} \, ds_4^2 + \frac{G_{\a\b}}{\sqrt{\S_1}} \,  (dy^\a + \A^\a+ \l_1 \hat \e^{\a\g} \B_\g) (dy^\b + \A^\b+ \l_1 \hat \e^{\b\g} \B_\g) \label{spflmet}
\ee 
where $\hat \e_{\a\b}$ is the $\e$ symbol ($\hat \e_{yt} = 1$) and $\S_1$ is given by

\be
\S_1 = (1+\l_1 \zeta)^2 + \l_1^2 e^{-2\phi} \det G_{\a\b}
\ee
The scalars $\zeta, \phi$ and the determinant $\det G_{\a\b}$ are inputs of the original geometry, which read
\be
\det G_{\a\b} = - \frac{1-2m f^{-1}}{H_1 H_5 } \;, \;\;\;\;\;e^{2\phi} = \frac{H_1}{H_5} \;, \;\;\;\;\; \zeta = \frac{2 m s_1 c_1}{f H_1}
\ee
Plugging in, we find that $\S_1$ takes the form

\be
\S_1=\frac{4 m^2 s_1^2 (s_1+c_1 \lambda_1)^2+ f^2 \left(1-\lambda_1^2\right)+2 f m \left(2 s_1^2+2 c_1 s_1 \lambda_1+\lambda_1^2\right)}{\left(f+2 m s_1^2\right)^2}
\ee
where the function $f$ is given in \eqref{deff}. Whenever $\l_1 \neq \pm 1$, the quantity $\S_1$ asymptotically approaches a constant. Nevertheless, when $\l_1 = \pm 1$, then $\S_1 \sim \mathcal{O}(r^{-2})$. This fact has a direct consequence on the asymptotic behaviour of the conformal factor $\Delta$, which under $T$ transforms as

\be
\Delta \r \Delta_1 = \S_1 \Delta
\ee
Thus, for $\l_1 = 1$, we can reduce the degree of divergence of $\Delta$ from $r^6$ to $r^4$. For precisely this value, $\S_1$ is

\be
\left. \S_1 \right|_{\l_1=1} = \frac{2m e^{2\d_1}}{f H_1} \;\;\; \Rightarrow \;\;\; \Delta_1  = 2 m e^{2\d_1} f^2 H_0 H_5
\ee
and $\Delta_1$ remains polynomial in $r$. The details of the above manipulations are given in appendix \ref{appttr}.

\bigskip

\noindent \emph{\textbf{The S transformation}}

\medskip

\noindent We can also act with the S transformation, whose action on the metric is very similar to \eqref{spflmet}
\be
ds^2_6 \r \sqrt{\S_2} \, ds_4^2 + \frac{G_{\a\b}}{\sqrt{\S_2}} \,  (dy^\a + \A^\a+ \l_2 \hat \e^{\a\g} \B_\g') (dy^\b + \A^\b+ \l_2 \hat \e^{\b\g} \B_\g')
\ee
The four-dimensional gauge field $\B_\a'$ is - roughly speaking - the  Hodge dual of  $\B_\a$. The quantity $\S_2$ is given by

\be
\S_2 = (1+ \l_2 \zeta')^2 + \l_2^2 \, e^{2\phi} \det G_{\a\b}
\ee
where the scalar $\zeta'$ is (roughly) the four-dimensional Hodge dual of the two form $\mathcal{C}_{\mu\nu}$. On the original black string background, 

\be
\zeta' = \frac{2 m s_5 c_5}{f H_5}
\ee
With these, we can compute $\S_2$ explicitly. It is given by

\be
\S_2=\frac{4 m^2 s_5^2 (s_5+c_5 \lambda_2)^2+ f^2 \left(1-\lambda_2^2\right)+2 f m \left(2 s_5^2+2 c_5 s_5 \lambda_2+\lambda_2^2\right)}{\left(f+2 m s_5^2\right)^2}
\ee
Note that, again, for $\l_2=1$ the asymptotics of $\S_2$ change from $\mathcal{O}(1)$ to $\mathcal{O}(r^{-2})$. The intermediate steps of this calculation can be found in appendix \ref{appstr}.

\bigskip

\noindent To summarize, the combined effect  of the S and T transformations on $\Delta$ is

\be
\Delta \r \S_1 \S_2 \Delta \label{trdelta}
\ee
When $\l_1=\l_2 =1$, the final value of $\Delta$ is

\be
\Delta_{fin} = 4 m^2 f H_0 \, e^{2\d_1+2\d_5}
\ee
This has the same large $r$ asymptotics as $\Delta_{sub}$ in \eqref{deltasub}, but it is not equal to it. Just like it is true of the subtracted geometries of the previous section, the uplifted black hole metric after an S and a T transformation becomes locally $AdS_3 \times S^3$. Consequently, there exists a coordinate transformation and a rescaling that takes it into the uplift of the subtracted geometry. We describe this transformation in the next subsection.

\subsection{The final geometry \label{secfingeo}}

Setting $\l_1=\l_2=1$, we find that the final metric is locally $AdS_3 \times S^3$

\be
ds_6^2 = ds_3^2 + \ell^2 \left[ d\th^2 + \sin^2 \th (d\phi+A^\phi)^2 + \cos^2 \th (d\psi+A^\psi)^2 \right] \label{fingeo}
\ee
where

\be
\ell^2 = 2 m e^{\d_1 + \d_5}
\ee
The three-dimensional Kaluza-Klein gauge fields are constant and read

\be
A^\phi = - (a d\tilde t + b d\tilde y) \;, \;\;\;\;\; A^\psi = - (a d\tilde y + b d\tilde t)
\ee
and the three-dimensional metric is

\be
ds_3^2 = \ell^2 \left[   r^2 (d\tilde y^2-d \tilde t^2) - (a^2 + b^2 -2m)d\tilde t^2-  2 a b d\tilde t d\tilde y + \frac{r^2 dr^2}{(r^2+a^2)(r^2+b^2)- 2 m r^2}  \right]
\ee
The new coordinates $\tilde t$ and $\tilde y$ are related to $t,y$ via

\be
\tilde t = \ell^{-2} (c_0 t - s_0 y) \;, \;\;\;\;\; \tilde y = \ell^{-2}(c_0 y - s_0 t)
\ee
Thus, the $\d_0$ dependence of the six-dimensional metric can be trivially undone via the above coordinate transformation. The geometry \eqref{fingeo} differs from the uplift of the subtracted geometry in two aspects: one needs to replace $\d_0$ by a new $\tilde \d_0$ and $\ell$ by $\tilde \ell$, given by

\be
\sinh \tilde \d_{0} = \frac{\Pi_s}{\sqrt{\Pi_c^2 - \Pi_s^2}} \;, \;\;\;\;\; \tilde \ell^2 = 2 m \sqrt{\Pi_c^2-\Pi_s^2}
\ee
This replacement amounts thus to a coordinate transformation and an overall rescaling. The metric can again be put in the form \eqref{btzform}, by defining 

\be
\rho = r^2 - m + \half (a^2+b^2) \;, \;\;\;\;\; u = y - t\;, \;\;\;\;\; v = y+t
\ee
The temperatures that we can read off are

\be
 T_\pm = \frac{ \sqrt{2 m -  (a\pm b)^2}}{2 \hat \ell^2} \, e^{\mp \tilde \d_0}
\ee
Pugging into Cardy's formula \eqref{cardyf}, we again get perfect match with the Bekenstein-Hawking entropy of the five-dimensional black hole which, in units of $G_5 = \pi/4$, reads

\be
S= 2 \pi m \sqrt{2m -(b-a)^2}\, (\Pi_c+\Pi_s) + 2 \pi m \sqrt{2m -(b+a)^2} \, (\Pi_c-\Pi_s)
\ee
Note that the coordinate transformation $\d_0 \r \tilde \d_0$ and the rescaling $\ell \r \tilde \ell$ were absolutely necessary in order to match the entropy in general. The only case in which these transformations are not needed is the neutral case $\d_i =0$, for which just the S and T transformations are enough to produce the subtracted geometry.

\section{Discussion}

In this paper, we have shown that all non-extremal four- and five-dimensional black holes with general rotation and charges can be ``untwisted'' to $AdS_3$ times a sphere, thus generalizing the work of \cite{skendy}. While it is possible that the untwisting may be done in several different ways \cite{skendy} - i.e. by using different choices of $SL(2,\mathbb{R})$ matrices - our particular choice is universal (it does not depend on any of the black hole parameters) and has a very simple duality interpretation. Moreover, the powerful solution generating techniques that we use allow us to easily construct solutions that interpolate between the original black holes and their subtracted geometries,  generalizing the work of \cite{deb}.

The most interesting application of our work would be to find the detailed microscopic interpretation of general non-extremal black holes. Using the duality interpretation of the Harrison and S,T transformations that we gave, both the four- and five-dimensional problems can be reduced to understanding the effect of the timelike Melvin twists, followed by several T-dualities, on the D0-D4 system. As usual, the five-dimensional case seems simpler, because only two  Melvin twists are required, rather than three. Given that the effect of a \emph{spacelike} Melvin twist on D4 branes is understood \cite{Bena:2002wg} - it simply corresponds to a mass deformation of the corresponding five-dimensional SYM theory - it is possible that also the timelike twist has a simple interpretation.

A different perspective on understanding the microscopic properties of general  non-extremal black holes is obtained by interpreting the backgrounds that interpolate between the uplifts of the original and subtracted geometries as  irrelevant deformations of the CFT dual to $AdS_3$. Infinitesimally, irrelevant deformations can be studied using the AdS/CFT correspondence \cite{vanRees:2011fr,vanRees:2011ir} and they can be  predictive in a certain regime of parameters \cite{deb}. Similar ideas have been promoted in \cite{kerrdip,Compere:2010uk} for understanding four and five-dimensional rotating \emph{extremal} black holes. In that case, the deformations are null and have a special structure that allows one to even study them at finite level \cite{nr}. The case of non-extremal black holes is complicated by the fact that the starting $AdS_3$  is at finite temperature and the deformation is not null \cite{deb}, but it would be interesting to further study and understand  these irrelevant deformations using the explicit interpolating solutions we have found . It would also be interesting to study various limits of the general black holes - e.g. the extremal limit and the connection with Kerr/CFT \cite{kerrcft} - and check whether the deformation simplifies in any case.

As we have discussed in sections \ref{dualint} and \ref{secfingeo}, the Harrison and timelike S and T transformations do not leave the entropy of the black holes invariant. It would be interesting to understand microscopically why this is the case, and whether there exist analogues of the transformations used in \cite{skendy} that do preserve the entropy. 
It would also be very interesting to better understand the thermodynamic properties of the ``twisted'' solutions, as well as the fate of the ergoregion of these black  holes, along the lines of \cite{Gibbons:2013yq}.

The Melvin twists and the irrelevant deformations are two complementary ways to understand the microscopic description of general asymptotically flat black holes in string theory, starting from the subtracted geometry. Nevertheless, it would be very interesting to understand how to adapt our use of the subtracted geometries to understand realistic black holes in realistic theories, such as pure Einstein or Einstein-Maxwell gravity. To this effect, the work of  \cite{Gibbons:2013yq}, who studied the action of Harrison transformations in pure Einstein-Maxwell theory, is of particular interest.

\bigskip
\noindent\textbf{Acknowledgements}

\medskip

\noindent We would like to thank Geoffrey Comp\`ere, Gary Gibbons, Juan Maldacena, Chris Pope and Balt van Rees for interesting discussions. Z.S. is grateful to Amitabh Virmani for discussions and for sharing his Mathematica notes. 
This work is  supported by the DOE grant DOE-EY-76-02-3071, and also by
 the Fay R. and Eugene L. Langberg Endowed Chair and the Slovenian
Research Agency (ARRS).

\appendix

\section{Useful formulae \label{usff}}

\noindent \emph{\textbf{The $3d\r 4d \r 5d$ lift}}

\medskip

\noindent Here we describe the relationship of  the four-dimensional fields that appear in the STU Lagrangian to the three-dimensional fields and dualized scalars, as well as  their uplift to five dimensions.

The four-dimensional  gauge fields can be reduced to three dimensions via

\be
A^\Lambda_{4d}= \zeta^\Lambda(dt+\omega_3) + A_3^\Lambda \label{uplifta}
\ee
Next, the three-dimensional gauge fields are dualized into scalars via

\be
-d\tilde{\zeta}_{\Lambda}  = e^{2U}(\mbox{Im} N)_{\Lambda \Sigma} \star_3 (d{A_3}^{\Sigma} + \zeta^{\Sigma} d\omega_3)+ (\mbox{Re} N)_{\Lambda \Sigma}d\zeta^{\Sigma} \non
\ee

\be
-d\sigma = 2 e^{4U} \star_3 d \omega_3 - \zeta^{\Lambda} d \tilde{\zeta}_{\Lambda} + \tilde{\zeta}_{\Lambda} d \zeta^{\Lambda} \label{3dd}
\ee
The relationship between the five-dimensional gauge fields and the four-dimensional ones is

\be
A_{5d}^I = - x^I (dz + A^0_{4d}) + A^I_{4d}
\ee
The real scalars in the five-dimensional $\N=2$ Lagrangian are given by

\be
h^I = f^{-1} y^I \;, \;\;\;\;\; f^3 = y^1 y^2 y^3
\ee
and the uplift of the metric is given in \eqref{5dlift}.

\bigskip

\noindent \emph{\textbf{The $5d \r 6d$ lift}}

\medskip

\noindent Here we describe the relationship between the five-dimensional black hole geometries and the six-dimensional black string ones that we use in section \ref{5dbhtw}. The reduction from six-dimensional Einstein frame to five dimensions is

\be
ds_6^2 = G_{yy} (dy+A^3_{5d})^2 + G_{yy}^{-\frac{1}{3}} ds_{5}^2 \;, \;\;\;\; G_{yy} = h_3^{-\frac{3}{2}}
\ee 
In terms of the four-dimensional fields that we have introduced in \eqref{6dgenmet}, we have

\be
A^3_{5d}=  \A^y + \frac{G_{yt}}{G_{yy}} (dt+ \A^t)
\ee
and

\be
ds_{5E}^2 = G_{yy}^{\frac{1}{3}}\, ds_4^2 + \frac{\det G_{\a\b}}{G_{yy}^{\frac{2}{3}}} (dt + \A^t)^2 \label{5deinst}
\ee
Comparing this expression with \eqref{5ddelta}, we find that

\be
\Delta = G_{yy} \left( \frac{f(f-2m)}{|\det G|}\right)^{\frac{3}{2}}
\ee
which is the equation we used to derive \eqref{trdelta}.

\section{The Harrison transformations as dualities}

\subsection{The $\a_0$ transformation \label{a0har}}

The action of the $\a_0$ Harrison transformation on the various three-dimensional 
fields in the theory can be read off from the transformation of the matrix $\mathcal{M}$ and reads

\be
e^{4U} \r \Xi_0^{-1}\, e^{4U} \;, \;\;\;\;\; y^I \r  \Xi_0^\half\, y^I \;, \;\;\;\;\; \Xi_0 = (1+\a_0 \zeta^0)^2 - \a_0^2\, f^{-3} e^{2U} \non
\ee

\be
\zeta^\Lambda \r \frac{\zeta^\Lambda (1+\a_0 \zeta^0) - \a_0\, x^\Lambda \, f^{-3} e^{2U} }{\Xi_0}\;, \;\;\;\;\; x^I \r x^I (1+\a_0 \zeta^0) - \a_0 \zeta^I \label{a0form1}
\ee
where we have introduced $x^0=1$. The transformation rules for $\tilde \zeta_\Lambda$ and $\s$ are rather cumbersome; instead, we can use \eqref{3dd} to compute the transformation of the Hodge dual fields $\om_3$, $ A_3^\Lambda$, which behave simply as

\be
\om_3 \r \om_3 - \a_0 A_3^0 \;, \;\;\;\;\; A_3^\Lambda \r A_3^\Lambda \label{a0form2}
\ee  
We would like to understand the effect of the $\a_0$ Harrison on the five-dimensional uplifted geometry. In terms of three-dimensional fields, the five-dimensional metric reads

\begin{equation}
 ds^2 = f^2( dz + \zeta_0( dt + \omega_3) + {A_3}^0 )^2 - f^{-1} e^{2U} ( dt + \omega_3)^2 + e^{-2U} f^{-1} {ds_3}^2 \label{5dmet3d}
\end{equation}
and the accompanying supporting gauge fields are

\be
A^I = - x^I (dz + \zeta^0(dt+ \om_3) + A_3^0) + \zeta^I (dt+\om_3) + A_3^I
\ee
Upon re-completing the squares in the required order, it is rather easy to see that the above transformations  are induced by a simple  change of coordinates 

\begin{equation}
 t \to t+ \alpha_0 z \label{cootr}
\end{equation}
in the five-dimensional background \eqref{5dmet3d}.

\subsection{The $\a_I$ transformations \label{aitransf}}

We will concentrate for concreteness on $\a_1$, which acts as

\be
e^{4U} \r \Xi_1^{-1} e^{4U} \;, \;\;\;\;\; y^1 \r  \Xi_1^\half\, y^1 \;, \;\;\;\;\; x^1 \r x^1 (1-\a_1 \tilde \zeta_1)-\a_1 \tilde \zeta_0 \non
\ee

\be
 \Xi_1 = (1-\a_1 \tilde \zeta_1)^2 - \a_1^2\, f^{-3} e^{2U} (x_2^2 + y_2^2)(x_3^2 + y_3^2) \label{a1form1}
\ee
Other fields that transform simply are

\bea
\tilde \zeta_0 & \r & \frac{\tilde \zeta_0 (1-\a_1 \tilde \zeta_1)-\a_1 x^1 e^{2U} f^{-3} (x_2^2 + y_2^2)(x_3^2 + y_3^2)}{\Xi_1}\non \\
\tilde \zeta_1 & \r & \frac{\tilde \zeta_1 (1-\a_1 \tilde \zeta_1)+\a_1 e^{2U} f^{-3} (x_2^2 + y_2^2)(x_3^2 + y_3^2)}{\Xi_1}\non \eea

\bea
 \zeta^2 & \r & \frac{ \zeta^2 (1-\a_1 \tilde \zeta_1)+\a_1 x^3 e^{2U} f^{-3} (x_2^2 + y_2^2)}{\Xi_1}
\non \\
 \zeta^3 & \r & \frac{\zeta^3 (1-\a_1 \tilde \zeta_1)+\a_1 x^2 e^{2U} f^{-3} (x_3^2 + y_3^2)}{\Xi_1} 
\label{a1form2}
\eea
The transformation rules for the remaining fields are rather complicated, and we will not include them here. 
The claim is that the above transformations are equivalent to four T-dualities along the $w^{3,4,5,6}$ directions, a coordinate transformation as in \eqref{cootr}, followed by four T-dualities back. 

Uplifting to ten dimensions, the type IIA string frame metric is

\be
ds_{10}^2 = ds_4^2 + y^1 (dw_1^2 + dw_2^2)+ y^2 (dw_3^2 + dw_4^2)+ y^3 (dw_5^2 + dw_6^2)
\ee
and the NS-NS B-field reads

\be
B^{(2)} =  - x^1 dw^1 \wedge dw^2 - x^2 dw^3 \wedge dw^4 - x^3  dw^5 \wedge dw^6
\ee
Under four T-dualities along $w^{3,4,5,6}$, the fields transform as

\be
 y^{2} \r \frac{y^2}{y_2^2 + x_2^2} \;, \;\;\;\;\;x^2 \r  \frac{x^2}{y_2^2+x_2^2} \label{tdx2y2}
\ee
and similarly for $x^3, y^3$. The action of the four T-dualities on the Ramond-Ramond fields is roughly to interchange $A^0_{4d}$ with (minus) the Hodge dual of $A^1$, and $A^2$ with $-A^3$. At the level of the three-dimensional scalars, we expect these exchanges to act as

\be
\zeta^0 \r - \tilde \zeta_1 \;, \;\;\;\; \zeta^1 \r \tilde \zeta_0 \;, \;\;\;\;\; \zeta^2 \r - \zeta^3 \label{zetarep}
\ee
while $U$ and $\om_3$ stay invariant. 

Back to the general formulae, it is easy to check that combining the replacements \eqref{tdx2y2}-\eqref{zetarep} with the coordinate transformation \eqref{a0form1}-\eqref{a0form2}, we obtain precisely the $\a_1$ Harrison transformation formulae \eqref{a1form1}-\eqref{a1form2}. Thus, for the subset of fields that we checked explicitly, this interpretation is correct.

\section{Explicit ``four-dimensional'' examples}

In this appendix we present explicit formulae for various four-dimensional black holes and interpolating geometries. The quotes above are due to the fact that we present the four-dimensional geometries either in terms of the three-dimensional scalar data, or in the form of the five-dimensional uplift.

In appendix \ref{gen4chbh} we present the scalar fields that yield the geometry of the general four-charge rotating black holes with three magnetic and one electric charge. To our best knowledge, the complete solution for all fields has not been published in the literature\footnote{The general solution with three magnetic charges can be found in \cite{vir}, whereas the explicit solution with two electric and two magnetic charges has been written down in \cite{CCLP}, minus the gauge potentials.}. In appendix \ref{deboer} we rederive the solution presented in \cite{deb} and relate our notation to theirs. Finally, in appendix \ref{intkerr} we present the five-dimensional uplift of the interpolating solution from the Kerr asymptotically flat black hole to its subtracted geometry. Since the formulae are rather cumbersome to write down, we present only the special cases $\a_1 =\a_2 =\a_3 =\a$ and $\a_2=\a_3=1$, both with $\a_0$ set to its subtracted value $\a_0 =0$.

\subsection{The general four-charge black hole \label{gen4chbh}}

We write herein the general four-dimensional asymptotically flat solution. The scalar field $U$ is given implicitly in \eqref{delta4ch}. Note that, despite the way it is presented, the expression for $\Delta$ is  completely symmetric under interchanging the charges. The other scalar fields are given by

\be
y^1 = \frac{\sqrt{\Delta}}{a^2 \cos^2\th + (r+ 2 m s_2^2)(r+ 2m s_3^2)}
\ee

\be
x^1 = \frac{2 a m \cos \th (c_0 c_1 s_2 s_3 - s_0 s_1 c_2 c_3)}{a^2 \cos^2\th + (r+ 2 m s_2^2)(r+ 2m s_3^2)}
\ee
The formulae for the remaining $x^i, y^i$ are obtained by permutations of the above. The next simplest scalar is

\be
\tilde \zeta_0 = \frac{2 a m \cos \th}{\Delta} \left[ s_0 c_1 c_2 c_3 (a^2 \cos^2 \th + r(r+ 2 m s_0^2)) - c_0 s_1 s_2 s_3(a^2 \cos^2 \th + (r-2m)(r+ 2m s_0^2))\right]
\ee
The formulae for the $\zeta^i$ are simply obtained from the above by replacing $\d_0 \leftrightarrow \d_i$. Next, we have

\bea
\zeta^0 & = & \frac{1}{\Delta} \bigl[  4 m^2 a^2 \cos^2\th \left((c_0^2+s_0^2) s_1 c_1 s_2 c_2 s_3 c_3 
- s_0 c_0 (2 s_1^2 s_2^2 s_3^2 + s_1^2 s_2^2+s_2^2 s_3^2 + s_3^2 s_1^2)\right) +\non \\
&& + 2 m s_0 c_0 \left(r a^2 \cos^2\th+\Pi_{i=1}^3 (r+ 2 m s_i^2) \right)   \bigr]
\eea
The expressions for the $\tilde \zeta_i$ are given by \emph{minus} the above expression, after replacing $\d_0 \leftrightarrow \d_i$. It may be useful to also note that $\tilde \zeta_{2,3}$ can also be written as

\be
\tilde \zeta_2 = \tilde \zeta^\star_2+ x^1 \zeta^3 \;, \;\;\;\;\;\tilde \zeta^\star_2= - \frac{2 m s_2 c_2 (r+ 2 m s_3^2)}{a^2 \cos^2\th + (r+ 2 m s_2^2)(r+ 2m s_3^2)} 
\ee
and similarly for $\tilde \zeta_3$, with the obvious replacements. Finally, the expression for $\s$ is given by

\be
\s = \frac{4 a m \cos \th (\Pi_c-\Pi_s)}{a^2 \cos^2\th + (r+ 2 m s_2^2)(r+ 2m s_3^2)} - (\zeta^0 \tilde \zeta_0 - \zeta^1 \tilde \zeta_1+ 2 x^1 \zeta^0 \tilde \zeta_1 + \zeta^2 \tilde \zeta^\star_2 + \zeta^3 \tilde \zeta^\star_3)
\ee
The fields $\s$ and $\tilde \zeta_\Lambda $ should be dualized to the one-forms $\om_3$ and $A^\Lambda_3$, which upon uplift yield the four-dimensional matter fields. It should be possible to check that
 $\om_3$ has the simple expression \eqref{eqom3}. 

\subsection{The static charged interpolating solution \label{deboer}}

The solution in the non-rotating charged case has been already given in \cite{deb}. We include a re-derivation of it in our notation. After four Harrison transformations with parameters $\a_A$, the scale factor $\Delta$ takes the form

\be
\Delta = \xi_0\, \xi_1 \,\xi_2\, \xi_3 
\ee
where
\be
\xi_A = (1-\a_A^2)\, r + \half m \, e^{2\d_A} \left(1+\a_A + e^{-2\d_A} (\a_A-1)\right)^2 \;, \;\;\;\; A \in \{0,\ldots, 3\}
\ee
The relationship between the parameters $\a_A$ and the ones - called $a_A$ - used to parametrize the interpolating solutions in \cite{deb} is

\be
a_A = \frac{\sqrt{1-\a_A^2}}{\sinh \d_A + \a_A \cosh \d_A}
\ee
Note that the values of the parameters $a_A$ which correspond to the subtracted geometry in \cite{deb} match precisely with the values quoted in \eqref{sola0}. One small advantage of our parametrization is that - unlike that of \cite{deb} - it is  not singular when one of the charges vanishes.

 The uplifted five-dimensional geometry takes the form

\be
ds_5^2 = (\xi_1 \xi_2 \xi_3)^{\frac{2}{3}} \left( d \Om_2^2 + \frac{dr^2}{G} \right) + (\xi_1 \xi_2 \xi_3)^{-\frac{1}{3}} ds_2^2
\ee
where

\be
ds_2^2= \xi_0 (dz+A^0)^2 - \frac{G}{\xi_0} dt^2  
\ee
and the Kaluza-Klein gauge field reads

\be
A^0 = \xi_0^{-1} \left( - \a_0 r + \half m (1+ e^{2\d_0}) \left(1+\a_0 + e^{-2\d_0} (\a_0-1)\right)  \right) \, dt
\ee
As before, the $\a_0$ dependence of the above metric, which is present only in the last parenthesis, $ds_2^2$, can be completely gauged away via the coordinate transformation \eqref{cootr}.  
The $3d$ Einstein metric reads

\be
ds_3^2 = (\xi_1 \xi_2 \xi_3)^2 \left( \frac{dr^2}{r(r-2m)} + \frac{ds_2^2}{\xi_1 \xi_2 \xi_3} \right)
\ee
When $\a_I=0$, this spacetime is $AdS_3$ of radius $\ell = 4 m e^{\frac{2}{3}(\d_1 + \d_2 + \d_3)}$. When at least two $\a_I$ are non-zero, including the asymptotically flat case, it is asymptotically conformal to $AdS_3$. It would be interesting if holography could be understood for this spacetime.

We also list the remaining five-dimensional fields, for completeness. We have

\be
h^I = \frac{\xi_I}{(\xi_1 \xi_2 \xi_3)^{\frac{1}{3}}} \;, \;\;\;\;\; A^I_{(5d)} = -\half m \left[ (1+\a_I)^2 e^{2\d_I}-(1-\a_I)^2 e^{-2\d_I} \right]\, \cos\th d\phi
\ee

\subsection{The neutral rotating interpolating solution \label{intkerr}}

While it is straightforward to generate the geometries that interpolate between the general charged rotating black holes and their subtracted geometry, the resulting formulae are rather uninspiring. Thus, we will limit ourselves to presenting only the simplest such rotating solution, for the neutral Kerr black hole. Introducing the notation

\be
\e_A = 1-\a_A^2
\ee
and

\be
\Pi_4 = \e_0 \e_1 \e_2 \e_3 \;, \;\;\;\;\; \Pi_3=\e_0 \e_1 \e_2 + \mbox{perms} 
\ee

\be
\Pi_2 = \e_0 \e_1 +\mbox{ perms} \;, \;\;\;\;\; \Pi_1 = \e_0 + \e_1+\e_2 + \e_3 
\ee
we find that the resulting warp factor is

\bea
\Delta & = & \Pi_4 \, r^4 + (2m \Pi_3 -8 m \Pi_4)\, r^3 + [4 m^2 \Pi_2 - 12 m^2 \Pi_3 + (24 m^2+ 2 a^2 \cos^2\th) \Pi_4 ]\, r^2 + \non \\ &+& 
 [8 m^3 \Pi_1 - 16 m^3 \Pi_2 + (24 m^3+ 2 a^2 m \cos^2\th) \Pi_3 - (32 m^3 + 8 a^2 m \cos^2\th) \Pi_4] r +\non \\
&+& 4m^2( a^2 \cos^2\th - 4 m^2) \Pi_1 + 16 m^4 \Pi_2 - 4 m^2 (4m^2 + a^2 \cos^2 \th) \Pi_3 +(a^2 \cos^2\th + 4 m^2)^2 \Pi_4+\non \\
&+& 16 m^4 + 8 a^2 m^2 \a_0 \a_1 \a_2 \a_3 \cos^2\th \label{expldel}
\eea
Note that it has all the properties that we have mentioned in section \ref{secsolgen}. Turning on $\e_{1,2,3}$ corresponds to turning on certain irrelevant deformations of the subtracted geometry. Since the solution for the remaining fields is still rather cumbersome, we will be focusing on two special cases:

\bi
\item equal deformations: $\e_1 = \e_2 =\e_3 =1-\a^2$
\item one nonzero deformation: $\e_1 = 1-\a_1^2$ and  $\e_2=\e_3 =0$
\ei
Since for neutral black holes we do not need to perform a $h_0$ Harrison transformation in order to reach the subtracted geometry, in both cases we will set $\a_0=0$.

\subsubsection*{Equal deformations}

In this subsection we present the uplifted five-dimensional geometry after a deformation with $\a_1 = \a_2 =\a_3 =\a$ and $\a_0=0$. As a useful intermediate step, we write the three-dimensional one-forms

\be
\om_3 = \frac{2 a  m r  \sin^2\th}{G} \, d \phi \;, \;\;\;\;\;  A_{3}^0 = \frac{2 a m (r-2m) \a^3 \sin^2 \th}{G} \, d \phi
\ee

\be
A_{3}^1 =A_{3}^2 =A_{3}^3 = - \frac{2 m X \a \cos\th}{G} \, d\phi
\ee
To write down the final interpolating solution, it is useful to introduce some shortcuts. Thus, we let

\be
 \rho = (1-\a^2)\, r + 2 m \a^2 \;, \;\;\;\;\; \e = 1-\a^2 \;, \;\;\;\;\; 
Y = \rho^2 + a^2 \e^2 \cos^2 \th 
\ee
but we still don't replace $\a$ by $\e$ when it appears with an odd power. The five-dimensional gauge fields then read

\be
A_{5d}^1 = -\frac{2m \a}{Y} [ a (\a dz -dt)  + (\rho^2 +a^2 \e^2 ) d\phi ]\, \cos \th \label{5dgf}
\ee
The components of the five-dimensional metric read

\be
g_{rr} = \frac{Y}{X} \;,\;\;\;\;\; g_{\th\th} = Y \;, \;\;\;\;\; g_{tz} = \frac{4 a^2 m^2 \a^3 \cos^2 \th}{Y^2}
\ee

\be
g_{t\phi} = - \frac{2 a m  \left(\rho ^3+a^2 \cos^2\th \epsilon ^2 (2 m (-1+\epsilon )+\rho )\right) \sin^2\th}{Y^2}
\ee

\bea
 g_{tt}& = & - \frac{1}{\e Y^2}\left[ \rho ^3 (-2 m+\rho )-2 a^2 \epsilon  \left(2 m^2 (-1+\epsilon )+m \epsilon  \rho -\epsilon  \rho ^2\right) \cos^2\theta+a^4 \epsilon ^4 \cos^4\theta \right] \non \\
g_{zz} & = & \frac{1 }{\e Y^2} \left[\rho ^3 (2 m (-1+\epsilon )+\rho )-2 a^2 \epsilon  \left(2 m^2 (-1+\epsilon )^2-m (-1+\epsilon ) \epsilon  \rho -\epsilon  \rho ^2\right) \cos^2 \theta+a^4 \epsilon ^4 \cos^4\theta \right] \non \\
g_{z\phi} &=& - \frac{2 a m \a^3 }{G Y^2 \e^2} \left[(2 m-\rho ) \rho ^3 (2 m (-1+\epsilon )+\rho )-2 a^2 \epsilon ^2 \left(4 m^3 (-1+\epsilon )+m^2 (6-4 \epsilon ) \rho + \right. \right.\non \\
&& \left. \left. +m (-3+\epsilon ) \rho ^2+\rho ^3\right) \cos^2\theta +a^4 \epsilon ^4 (2 m-\rho ) \cos^4\theta \right]\,\sin^2\th\non \\
g_{\phi\phi} & = & \frac{\sin^2\th}{Y^2} \left[ \rho ^3 \left(\rho ^3+a^2 \epsilon ^2 (-2 m (-2+\epsilon )+\rho )\right)+2 a^2 \epsilon ^2 \left(\rho ^3 (m (-2+\epsilon )+\rho )+a^2 \epsilon ^2 \left(2 m^2 (-1+\epsilon )-\right. \right.\right. \non \\
&- & \left. \left.\left.  m (-2+\epsilon ) \rho +\rho ^2\right)\right) \cos^2 \th +a^4 \epsilon ^4 \left(-4 m^2 (-1+\epsilon )+a^2 \epsilon ^2+2 m (-2+\epsilon ) \rho +\rho ^2\right) \cos^4\th \right]
\eea
We were unable to find much structure in the above solution, but it would be interesting if it existed. To reduce to three dimensions, we write the metric as

\be
ds_5^2 = e^{-2U -2V} ds_3^2 + 4m^2 e^{2U} d\th^2 + 4 m^2 e^{2V} \sin^2\th (d\phi + \hat A)^2 \label{5dto3dred}
\ee
and find that, to first order in $\e$,

\be
e^{2U}= e^{2V} =  1 + \frac{(r-2m) \e}{m}+ \ldots
\ee

\be
\hat A_t = - \frac{a}{4 m^2} + \frac{3 a (r-2m) \e}{8 m^3} + \ldots \;, \;\;\;\;\; \hat A_z = \frac{a}{4 m^2} + \frac{3 a (m-r) \e}{8 m^3} + \ldots 
\ee
The $r$-dependence of the three-dimensional vector field indicates the presence of a $(1,2)$ operator (in addition to the $(2,2)$ ones found in \cite{deb}), whose coupling is proportional to the rotation parameter $a$. Note that unlike in the static case, beyond the leading order in $\e$, $U$ and $V$ will no longer be equal. The deformation of the three-dimensional Einstein metric reads

\be
g_{tt} = - \frac{2 m r - 4 m^2 + a^2}{4 m^2} - \frac{3(r-2m)^2 \e}{4m^2} \;, \;\;\;\;\; \;\; g_{zz} = \frac{2 m r-a^2}{4m^2} + \frac{3(r^2 - 2 m r +a^2)\e}{4 m^2}  \non
\ee

\be
g_{tz} = \frac{a^2}{4 m^2} - \frac{3 a^2 \e}{8 m^2} \;, \;\;\;\;\; g_{rr} = \frac{4 m^2}{X} + \frac{12 m (r-2m) \e}{X}
\ee
There are also additional massive vector fields coming from the dimensional reduction of the five-dimensional gauge field \eqref{5dgf}.

\bigskip

\subsubsection*{Single deformation}

To study the effect of a single deformation, we set $\a_2=\a_3=1$. The angular velocity $\om_3$ stays the same, whereas the gauge fields change to

\be
A_3^0 = \frac{2 a m (r-2m) \a_1 \sin^2\th}{G} \;, \;\;\;\; A_3^1 = - \frac{2 m X \a_1 \cos \th}{G} \;, \;\;\;\;\; A_3^2=A_3^3 = - \frac{2 m X \cos \th}{G}
\ee
The five-dimensional gauge fields  read

\be
A^1_{5d} =- \left( 2 m \a_1 d\phi + \frac{a(dz-\a_1 dt)}{2m}\right) \cos \th
\ee

\be
A^2_{5d} = A^3_{5d} =  \left( - 2 m  d\phi + \frac{a(dt-\a_1 dz)}{\rho_1}\right) \cos \th
\ee
where we have again introduced the shorthand

\be
\rho_1 = (1-\a_1^2) \, r + 2 m \a_1^2 
\ee
Note that the magnetic flux through the sphere is decreased. The metric takes the form \eqref{5dto3dred} with $U=V$, where

\be
e^{2U} = \left(\frac{\rho_1}{2m}\right)^{\frac{2}{3}}\;, \;\;\;\;\;\hat A = \frac{a (\a_1 dz -dt)}{2 m \rho_1} 
\ee

\be
g_{tt} = - \frac{r^2 (1-\a_1^2) + 2 m r (2 \a_1^2-1) +a^2 - 4m^2 \a_1^2}{4m^2} \;, \;\;\;\;\; g_{tz} = \frac{a^2 \a_1}{4m^2}\non
\ee

\be
g_{zz} = \frac{r^2(1-\a_1^2)+ 2 m r \a_1^2 -a^2 \a_1^2}{4m^2}\;, \;\;\;\;\; g_{rr} = \frac{\rho_1^2}{X}
\ee
The scalars that support the geometry are

\be
h^1 = \left( \frac{\rho_1}{2m}\right)^{\frac{2}{3}} \;, \;\;\;\;\;h^2=h^3 = \left( \frac{2m}{\rho_1}\right)^{\frac{1}{3}}
\ee
It would be interesting if one could construct a consistent truncation of the five-dimensional action to three-dimensions, that  contains this solution and then  perform a detailed  holographic analysis. 

\section{Details of the spectral flows \label{5dbhint}}

\subsection{The general black string solution \label{genbs}}

The Einstein-frame metric of the general six-dimensional black string solution \cite{Cvetic:1998xh} is given by

\be
ds_6^2 = ds_4^2 + G_{\a\b} (dy^\a + \A^\a) (dy^\b + \A^b) 
\ee
where $y^\a = \{y,t\}$ and

\bea
ds_4^2 &= &\sqrt{H_1 H_5} \left[\left(a^2 + r^2 + \frac{2 a^2 m  \sin^2\th}{f - 2 m } \right) \sin^2 \th d\phi^2 + \left( b^2 + r^2 + \frac{2 b^2 m  \cos^2\th}{f - 2 m }\right) \cos^2\th d\psi^2+ \right.\non \\
&&\left. +\frac{2 a b m  \sin^2\th \cos^2\th}{f - 2 m } \,2 d\phi d\psi + \frac{f r^2 dr^2}{ (r^2+a^2)(r^2+b^2)-2 m r^2 }  + f d\th^2
  \right] \label{4dmetstr}
\eea
The four-dimensional base metric is related to $d\hat s^2_4$ that appears in \eqref{5ddelta} by the rescaling

\be
ds_4^2 = f \sqrt{H_1 H_5} \, d\hat s_4^2 \label{dshatsq}
\ee
The Kaluza-Klein gauge fields read

\be
\A^y  =  2 m  \left(a \frac{ s_0 c_1 c_5}{f - 2 m } - b \frac{c_0 s_1 s_5}{f} \right) \sin^2 \th d\phi +  2 m  \left(b \frac{ s_0 c_1 c_5}{f - 2 m } - a \frac{c_0 s_1 s_5}{f} \right) \cos^2 \th d\psi \non \\
\ee

\be
\A^t  =  2 m  \left(a \frac{c_0 c_1 c_5}{f - 2 m } - b \frac{s_0 s_1 s_5}{f} \right) \sin^2 \th d\phi  + 2 m  \left(b \frac{ c_0 c_1 c_5}{f - 2 m } - a \frac{s_0 s_1 s_5}{f} \right) \cos^2 \th d\psi 
\ee

\be
G_{\a\b} = \frac{1}{\sqrt{H_1 H_5}}\left(\begin{array}{cc} H_0 & - \frac{m  \sinh 2 \d_0}{f} \\ - \frac{m  \sinh 2 \d_0}{f} & \frac{2 m  \cosh^2\d_0}{f}-1  \end{array}\right)\;, \;\;\;\;\; \det G_{\a\b} = - \frac{1-2m f^{-1}}{H_1 H_5 } \label{gab}
\ee
We have defined

\be
H_i = 1+ \frac{2m  \sinh^2\d_i}{f} \;, \;\;\;\;\; f = r^2 + a^2 \cos^2\th + b^2 \sin^2\th
\ee
The solution is also supported by the ten-dimensional dilaton, which in RR frame reads

\be
e^{2\phi} = \frac{H_1}{H_5}
\ee
and by the Ramond-Ramond two-form field, which can be found in \cite{Giusto:2004id}. Using \eqref{decc}, we  also decompose the six-dimensional $C^{(2)}$ field to four dimensions, obtaining

\be
\zeta = \frac{2 m s_1 c_1}{f H_1} \;, \;\;\;\;\; \C = m s_5 c_5 \cos^2\th \left( \frac{a^2 + r^2}{f} + \frac{a^2 + r^2 -2m}{f-2m}\right) d\phi\wedge d\psi
\ee

\be
\B_y  =  2 m  \left(a \frac{ c_0 s_1 c_5}{f - 2 m } - b \frac{s_0 c_1 s_5}{f} \right) \sin^2 \th d\phi +  2 m  \left(b \frac{ c_0 s_1 c_5}{f - 2 m } - a \frac{s_0 c_1 s_5}{f} \right) \cos^2 \th d\psi \non \\
\ee

\be
\B_t  =  2 m  \left(-a \frac{s_0 s_1 c_5}{f - 2 m } + b \frac{c_0 c_1 s_5}{f} \right) \sin^2 \th d\phi  + 2 m  \left(-b \frac{ s_0 s_1 c_5}{f - 2 m } + a \frac{c_0 c_1 s_5}{f} \right) \cos^2 \th d\psi 
\ee
For our future manipulations, it is useful to introduce the scalar  $\zeta'$, defined as

\be
d \zeta' = v^2 \sqrt{|\det G|} \star_4 H^{(3)} \;, \;\;\;\;\; H^{(3)} = d \,\C - \half \A^\a \wedge d\B_\a - \half \B_\a \wedge d \A^\a
\ee
where $v^2= e^{2\phi}$ is the volume of the internal four-torus. We simply find 

\be
\zeta' = \frac{2 m s_5 c_5}{f H_5}
\ee

\subsection{The T transformation \label{appttr}}

The action of the T transformation on the four-dimensional fields is

\be
e^{2\phi_1} = e^{2\phi} \S_1 \;, \;\;\;\;\; \zeta_1 = \frac{\zeta + \l_1(\zeta^2 + e^{-2\phi} \det G)}{\S_1}\;, \;\;\;\;\; \A^\a_1 = \A^\a+\l_1 \hat\e^{\a\b} \B_\b
\ee
where $\S_1$ is given by

\be
\S_1 = (1+\l_1 \zeta)^2 + \l_1^2 e^{-2\phi} \det G
\ee
Note that in Lorentzian signature, the transformation of $\A^\a$ differs by a sign from its spacelike counterpart, due to the different definition of $\hat \e^{\a\b}$. The fields $\B_\a$ and $\zeta'$ are unchanged, and

\be
G_{\a\b}^1 = \frac{G_{\a\b}}{\sqrt{\S_1}} \;, \;\;\;\;\; {ds_4^1}^2 = \sqrt{\S_1}\, ds_4^2
\ee

\subsection{The S transformation \label{appstr}}

Let us now perform the S transformation. Its action on the four-dimensional fields is 

\be
e^{2\phi_2} = \S_2^{-1} e^{2\phi_1} = \frac{\S_1}{\S_2}\, e^{2\phi} \;, \;\;\;\;\; \zeta_2 = \zeta_1\;, \;\;\;\;\; \A_2^\a = \A_1^\a+\l_2 \hat \e^{\a\b} \B'_\b 
\ee
The factor $\S_2$ is given by

\be
\S_2 = (1+ \l_2 \zeta')^2 + \l_2^2 \, e^{2\phi_1} \det G_1 = (1+ \l_2 \zeta')^2 + \l_2^2 \, e^{2\phi} \det G
\ee
The four-dimensional gauge field $\B'_\a$ is determined by

\be
d \B'_\a = e^{2\phi_1} \e_\a{}^\b \star_4 d\B_\b  + \zeta' \hat \e_{\a\b} \, d\A^\b_1 - \frac{e^{2\phi_1} \zeta_1}{\sqrt{\det G}} G_{\a\b} \star_4 d\A^\b_1
\ee
Note the sign difference with respect to the Euclidean signature formulae in \cite{stubena}. To solve for $\B'_\a$, it is useful to rewrite the above expression in terms of the fields before the T transformation

\bea
d\B'_\a & = &e^{2\phi} \,\frac{G_{\a\b}}{\sqrt{\det G}} \left( \hat \e^{\b\g} \star_4 d\B_\g - \zeta \star_4 d\A^\b \right) + \zeta' \, \hat \e_{\a\b}\, d \A^b + \\
&+&  \l_1 \left[ e^{2\phi} \,\frac{G_{\a\b}}{\sqrt{\det G}} \left( \zeta\, \hat \e^{\b\g} \star_4 d\B_\g - (\zeta^2 + e^{-2\phi} \det G )\, \star_4 d\A^\b \right)+ \zeta' \, d\B_\a\right]\non
\eea
Writing

\be
\B'_\a = \B^{'(0)}_\a+ \l_1 \B^{'(1)}_\a
\ee
and integrating the above equation, we find

\bea
B_y^{'(0)} &=& 2m \left( \frac{a c_0 c_1 s_5}{f-2m}-\frac{b s_0 s_1 c_5}{f}\right) \sin^2 \th d\phi-2m \left( \frac{a s_0 s_1 c_5}{f}-\frac{b c_0 c_1 s_5}{f-2m} \right) \cos^2 \th d\psi \non \\
B_t^{'(0)} &=& -2m \left( \frac{a s_0 c_1 s_5}{f-2m}-\frac{b c_0 s_1 c_5}{f}\right) \sin^2 \th d\phi+2m \left( \frac{a c_0 s_1 c_5}{f}-\frac{b s_0 c_1 s_5}{f-2m} \right) \cos^2 \th d\psi \non \\ &&
\eea
and

\bea
B_y^{'(1)} &=& 2m \left( \frac{a c_0 s_1 s_5}{f-2m} -\frac{b s_0 c_1 c_5}{f}\right) \sin^2 \th d\phi-2m \left( \frac{a s_0 c_1 c_5}{f}-\frac{b c_0 s_1 s_5}{f-2m} \right) \cos^2 \th d\psi \non \\
B_t^{'(1)} &=& - 2m \left(\frac{a s_0 s_1 s_5}{f-2m}-\frac{b c_0 c_1 c_5}{f}\right) \sin^2 \th d\phi+2m \left( \frac{a c_0 c_1 c_5}{f}-\frac{b s_0 s_1 s_5}{f-2m} \right) \cos^2 \th d\psi \non \\ &&
\eea
Finally, the metric becomes

\be
G_{\a\b}^2 = \frac{G_{\a\b}^1}{\sqrt{\S_2}} = \frac{G_{\a\b}}{\sqrt{\S_1 \S_2}} \;, \;\;\;\;\;\; {ds_4^2}^2 = \sqrt{\S_2} \, {ds_4^1}^2 = \sqrt{\S_1 \S_2}\, ds_4^2
\ee


\begin{thebibliography}{99} 
 

\bibitem{CYI}
  M.~Cveti\v c and D.~Youm,
 \emph{``Entropy of nonextreme charged rotating black holes in string theory''},
  Phys.\ Rev.\  D {\bf 54}, 2612 (1996),
  arXiv: hep-th/9603147.

\bibitem{CYII}
  M.~Cveti\v c and D.~Youm,
\emph{ ``General rotating five-dimensional black holes of toroidally compactified
  heterotic string,''}
  Nucl.\ Phys.\  B {\bf 476}, 118 (1996), 
  arXiv: hep-th/9603100.



\bibitem{Larsen}
 F.~Larsen,
  \emph{``A string model of black hole microstates,''}
  Phys.\ Rev.\  D {\bf 56}, 1005 (1997),
  arXiv: hep-th/9702153.
  
  
\bibitem{CL97I}
 M.~Cveti\v c and F.~Larsen,
 \emph{ ``General rotating black holes in string theory: Grey body factors and event
  horizons,''}
  Phys.\ Rev.\  D {\bf 56}, 4994 (1997),
  arXiv: hep-th/9705192.

\bibitem{CL97II}
M.~Cveti\v c and F.~Larsen,
 \emph{``Grey body factors for rotating black holes in four-dimensions,''}
  Nucl.\ Phys.\  B {\bf 506}, 107 (1997),
  arXiv: hep-th/9706071.

\bibitem{Castro:2010fd}
  A.~Castro, A.~Maloney and A.~Strominger,
 \emph{ ``Hidden Conformal Symmetry of the Kerr Black Hole,''}
  Phys.\ Rev.\ D {\bf 82} (2010) 024008, 
  arXiv: 1004.0996 [hep-th].


\bibitem{CL11I}
  M.~Cveti\v c and F.~Larsen,
\emph{ ``Conformal Symmetry for General Black Holes,''}
  JHEP {\bf 1202}, 122 (2012),
  arXiv: 1106.3341 [hep-th].


\bibitem{CL11II}
M. Cveti\v c and F. Larsen,
\emph{``Conformal Symmetry for Black Holes in Four Dimensions,''}
 JHEP {\bf 1209}, 076 (2012), 
  arXiv: 1112.4846 [hep-th].

 
   \bibitem{conic} 
  M.~Cveti\v  c and G.~W.~Gibbons,
 \emph{ ``Conformal Symmetry of a Black Hole as a Scaling Limit: A Black Hole in an Asymptotically Conical Box,''}
  JHEP {\bf 1207}, 014 (2012),
  arXiv: 1201.0601 [hep-th].
 
  
\bibitem{Duff:1995sm} 
  M.~J.~Duff, J.~T.~Liu and J.~Rahmfeld,
  \emph{``Four-dimensional string-string-string triality,''}
  Nucl.\ Phys.\ B {\bf 459}, 125 (1996), arXiv:
  hep-th/9508094.






  
\bibitem{deb}
  M.~Baggio, J.~de Boer, J.~I.~Jottar and D.~R.~Mayerson,
 \emph{``Conformal Symmetry for Black Holes in Four Dimensions and Irrelevant Deformations,''}
  arXiv: 1210.7695 [hep-th].



 \bibitem{Chakraborty:2012nu}
  A.~Chakraborty and C.~Krishnan,
\emph{``Subttractors,''}
  arXiv: 1212.1875 [hep-th].
  
\bibitem{Chakraborty:2012fx} 
  A.~Chakraborty and C.~Krishnan, \emph{
  ``Attraction, with Boundaries,''}
  arXiv: 1212.6919 [hep-th].

\bibitem{berggan} A.~Bergman and O.~Ganor, \emph{``Dipoles, twists and noncommutative gauge theory''},
JHEP {\bf 0010} (2000) 018, arXiv: hep-th/0008030.

\bibitem{dasgupta} 
  K.~Dasgupta and M.~M.~Sheikh-Jabbari,
  \emph{``Noncommutative dipole field theories''},
  JHEP {\bf 0202}, 002 (2002),
  arXiv: hep-th/0112064.

\bibitem{Bergman:2001rw} 
  A.~Bergman, K.~Dasgupta, O.~J.~Ganor, J.~L.~Karczmarek and G.~Rajesh,
  \emph{``Nonlocal field theories and their gravity duals''}, 
  Phys.\ Rev.\ D {\bf 65}, 066005 (2002), 
  arXiv: hep-th/0103090.


\bibitem{stromdipwei}
  W.~Song and A.~Strominger,
  \emph{``Warped AdS3/Dipole-CFT Duality,''}
  arXiv: 1109.0544 [hep-th].



\bibitem{kerrdip}
  S.~El-Showk and M.~Guica,
  \emph{``Kerr/CFT, dipole theories and nonrelativistic CFTs,''}
  arXiv: 1108.6091 [hep-th].


\bibitem{kerrcft} M.~Guica, T.~Hartman, W.~Song, A.~Strominger, \emph{``The Kerr/CFT correspondence''}, 
Phys.Rev. {\bf D80} (2009) 124008,  arXiv: 0809.4266 [hep-th].

\bibitem{nr} 
  M.~Guica, K.~Skenderis, M.~Taylor and B.~C.~van Rees,
  \emph{``Holography for Schrodinger backgrounds,''}
  JHEP {\bf 1102}, 056 (2011),
  arXiv: 1008.1991 [hep-th].  
  
  \bibitem{vir}
A.~Virmani,
 \emph{``Subtracted Geometry From Harrison Transformations,''}
 JHEP {\bf 1207}, 086 (2012), 
arXiv: 1203.5088 [hep-th].

\bibitem{Gibbons:2013yq}
  G.~W.~Gibbons, A.~H.~Mujtaba and C.~N.~Pope, \emph{
  ``Ergoregions in Magnetised Black Hole Spacetimes,''}
  arXiv: 1301.3927 [gr-qc].

 
\bibitem{Yazadjiev:2013hxa}
  S.~S.~Yazadjiev,\emph{
  ``Electrically charged dilaton black holes in external magnetic field,''}
  arXiv: 1302.5530 [gr-qc]. 




\bibitem{Bena:2008wt}
  I.~Bena, N.~Bobev and N.~P.~Warner,
 \emph{ ``Spectral Flow, and the Spectrum of Multi-Center Solutions,''}
  Phys.\ Rev.\ D {\bf 77}, 125025 (2008)
  arXiv: 0803.1203 [hep-th].



\bibitem{skendy}
  K.~Sfetsos and K.~Skenderis, \emph{
  ``Microscopic derivation of the Bekenstein-Hawking entropy formula for nonextremal black holes,''}
  Nucl.\ Phys.\ B {\bf 517} (1998) 179, arXiv:
  hep-th/9711138.

  
\bibitem{Hyun:1997jv} 
  S.~Hyun,
  \emph{``U duality between three-dimensional and higher dimensional black holes,''}
  J.\ Korean Phys.\ Soc.\  {\bf 33}, S532 (1998), arXiv:
  hep-th/9704005.

\bibitem{Boonstra:1997dy} 
  H.~J.~Boonstra, B.~Peeters and K.~Skenderis, \emph{
  ``Duality and asymptotic geometries,''}
  Phys.\ Lett.\ B {\bf 411}, 59 (1997), 
 arXiv: hep-th/9706192.


  \bibitem{stubena} 
  I.~Bena, M.~Guica and W.~Song,
  \emph{``Un-twisting the NHEK with spectral flows,''}
  arXiv: 1203.4227 [hep-th].


  \bibitem{CCLP}
  Z.~W.~Chong, M.~Cveti\v c, H.~L\"u and C.~N.~Pope,
 \emph{``Charged rotating black holes in four-dimensional gauged and ungauged
 supergravities,''}
  Nucl.\ Phys.\  B {\bf 717}, 246 (2005), 
  arXiv: hep-th/0411045.

\bibitem{Dong:2012se}
  X.~Dong, S.~Harrison, S.~Kachru, G.~Torroba and H.~Wang, \emph{
  ``Aspects of holography for theories with hyperscaling violation,''}
  JHEP {\bf 1206} (2012) 041, 
  arXiv: 1201.1905 [hep-th].

\bibitem{Sen:1994wr} 
  A.~Sen, \emph{
  ``Strong - weak coupling duality in three-dimensional string theory,''}
  Nucl.\ Phys.\ B {\bf 434}, 179 (1995), arXiv:
  hep-th/9408083.


\bibitem{Bossard:2009we} 
  G.~Bossard, Y.~Michel and B.~Pioline,\emph{
  ``Extremal black holes, nilpotent orbits and the true fake superpotential,''}
  JHEP {\bf 1001}, 038 (2010),
  arXiv: 0908.1742 [hep-th].

 \bibitem{klemm}
  S.~Bertini, S.~L.~Cacciatori and D.~Klemm,
 \emph{``Conformal structure of the Schwarzschild black hole,''}
  arXiv: 1106.0999 [hep-th].

\bibitem{Brown:1986nw} 
  J.~D.~Brown and M.~Henneaux,
  \emph{``Central Charges in the Canonical Realization of Asymptotic Symmetries: An Example from Three-Dimensional Gravity,''}
  Commun.\ Math.\ Phys.\  {\bf 104}, 207 (1986).

\bibitem{Papadopoulos:1996uq} 
  G.~Papadopoulos and P.~K.~Townsend, \emph{``Intersecting M-branes,''}
  Phys.\ Lett.\ B {\bf 380}, 273 (1996).

\bibitem{Tseytlin:1996bh} 
  A.~A.~Tseytlin, \emph{
  ``Harmonic superpositions of M-branes,''}
  Nucl.\ Phys.\ B {\bf 475}, 149 (1996), arXiv:
  hep-th/9604035.


\bibitem{msw}  J.~M.~Maldacena, A.~Strominger and E.~Witten,
  \emph{``Black hole entropy in M theory,''}
  JHEP {\bf 9712} (1997) 002, arXiv:
  hep-th/9711053.



\bibitem{Cvetic:1998xh}
  M.~Cvetic and F.~Larsen, \emph{
  ``Near horizon geometry of rotating black holes in five-dimensions,''}
  Nucl.\ Phys.\ B {\bf 531} (1998) 239, arXiv:
  hep-th/9805097.


\bibitem{Bena:2002wg} 
  I.~Bena and C.~Ciocarlie,\emph{
 ``Exact N=2 supergravity solutions with polarized branes,''}
  Phys.\ Rev.\ D {\bf 70}, 086005 (2004), arXiv:
  hep-th/0212252.





\bibitem{vanRees:2011fr}
  B.~C.~van Rees, \emph{``Holographic renormalization for irrelevant operators and multi-trace counterterms,''}
  JHEP {\bf 1108} (2011) 093,
  arXiv: 1102.2239 [hep-th].

\bibitem{vanRees:2011ir} 
  B.~C.~van Rees, \emph{``Irrelevant deformations and the holographic Callan-Symanzik equation,''}
  JHEP {\bf 1110}, 067 (2011),
  arXiv: 1105.5396 [hep-th].


\bibitem{Compere:2010uk} 
  G.~Compere, W.~Song and A.~Virmani, \emph{
  ``Microscopics of Extremal Kerr from Spinning M5 Branes,''}
  JHEP {\bf 1110}, 087 (2011), 
  arXiv: 1010.0685 [hep-th].


\bibitem{Giusto:2004id} 
  S.~Giusto, S.~D.~Mathur and A.~Saxena,
\emph{``Dual geometries for a set of 3-charge microstates,''}
  Nucl.\ Phys.\ B {\bf 701}, 357 (2004), arXiv:
  hep-th/0405017.

\end{thebibliography}
\end{document}